\documentclass[%
 aip,
 amsmath,amssymb,
 reprint,%
]{revtex4-1}

\usepackage{graphicx}
\usepackage{dcolumn}
\usepackage{bm}

\usepackage{amssymb}
\usepackage[font=scriptsize]{caption}
\usepackage{amsmath}
\usepackage{bbold}
\usepackage{etoolbox}
\usepackage{multirow, booktabs}
\usepackage{tabularx}
\usepackage{braket}
\usepackage{float}
\usepackage[caption=false]{subfig}
\usepackage[utf8]{inputenc}
\usepackage[T1]{fontenc}
\usepackage{mathptmx}

\begin{document}

\title[Geometric band properties in strained ML TMDs]{Geometric band properties in strained 
monolayer transition metal dichalcogenides using simple band structures\\}

\author{Shahnaz Aas}

\author{Ceyhun Bulutay}
 \email{bulutay@bilkent.edu.tr}
\affiliation{ 
Department of Physics, Bilkent University, 06800, Bilkent, Ankara, Turkey
}%

\date{\today}

\begin{abstract}
Monolayer transition metal dichalcogenides (TMDs) bare large Berry curvature hotspots readily exploitable for geometric band effects.
Tailoring and enhancement of these features via strain is an active research direction. Here, we consider
spinless two- and three-band, and spinful four-band models capable to quantify Berry curvature and orbital magnetic moment of 
strained TMDs. First, we provide a $k\cdot p$ parameter set for MoS$_2$, MoSe$_2$, WS$_2$, and WSe$_2$ in the light of the 
recently released ab initio and experimental band properties. Its validity range extends from $K$ valley edge to about hundred 
millielectron volts into valence 
and conduction bands for these TMDs. To expand this over a larger part of the Brillouin zone, we incorporate strain to
an available three-band tight-binding Hamiltonian. With these techniques we demonstrate that both the Berry curvature 
and the orbital magnetic moment can be doubled compared to their intrinsic values by applying typically a 2.5\% biaxial 
tensile strain. These simple band structure tools can find application in the quantitative device modeling of the geometric 
band effects in strained monolayer TMDs.
\end{abstract}

\maketitle

\section{\label{sec:intro}Introduction\protect\\}
The monolayer transition metal dichalcogenides (TMDs) of the semiconducting $2H$ polytype avail wide range of electrical, magnetic, 
optical, and mechanical control and tunability. \cite{xu_2014,manzeli_2017,mak_2018} Their valley-contrasting properties associated 
with the so-called inequivalent $K$ valleys at the corners of hexagonal Brillouin zone grant information carriers the opportunity 
to experience non-dissipative electronics. \cite{yamamoto_2015} Unlike the similar multiple conduction band valleys in conventional 
bulk silicon electronics, in TMDs the valley  degree of freedom is practically an individually accessible quantum label. 
\cite{schaibley_2016} For instance, in the so-called valley Hall effect an in-plane electric field initiates a valley current 
in the transverse in-plane direction, \cite{xiao_2007,yao_2008,xiao_2012} which has been confirmed by both optical 
\cite{mak_2014} and transport \cite{wu_2019} measurements. 

At the heart of these valley-based physics there lies the sublattice-driven orbital angular momentum. \cite{yao_2008,cao_2012} 
Alternatively, from the perspective of quantum geometrical band properties, \cite{vanderbilt_2018} the foregoing effects can be 
attributed to the Berry curvature (BC) and orbital magnetic moment (OMM). \cite{xiao_2010} Both of them take part in 
various phenomena such as the dichroic selection rules in optical absorption, \cite{zhang_2018,cao_2018} or the excitonic 
$p$ level energy splitting which is proportional to the BC flux; \cite{srivastava_2015a,zhou_2015}
OMM accounts for the interatomic currents (self-rotating motion of the electron wavepacket) \cite{chang_2008} responsible for the valley 
$g$-factor in TMDs. \cite{brooks_2017,rybkovskiy_2017} Thus, by breaking time-reversal symmetry with a perpendicular magnetic field, 
a valley Zeeman splitting is introduced in addition to the well-known spin Zeeman effect. \cite{aivazian_2015,srivastava_2015b}
Very recently, through their intimate connection with the orbital angular momentum, these geometric band properties are locally mapped 
in momentum space using circular dichroism angle-resolved photoelectron spectroscopy. \cite{schuller_2019} 

A unique advantage of TMDs is their mechanical deformability up to at least 10\% in their lattice constants without degradation.\cite{bertolazzi_2011}
Undoubtedly it is bound to have ramifications on the quantum geometric band properties, where a quantification inevitably
necessitates band structure tools reliable under strain.
The $k\cdot p$ method has been the first resort because of its simplicity, starting with graphene\cite{xiao_2007} and carried over to other
two-dimensional materials. \cite{kormanyos_2015,rostami_2015,pearce_2016,sevik_2017,rybkovskiy_2017}
Very recently a strained parametrization is also offered, \cite{fang_2018} which we used to successfully explain the experimental photoluminescence 
peak shifts in strained TMDs. \cite{aas_2018} On the other hand, it has a number of shortcomings especially for studying carrier 
transport away from the $K$ point. Namely, it is isotropic, preserves the electron-hole symmetry, and remains parabolic. 
In contrast, TMDs possess the trigonal warping (TW) of the isoenergy contours which leads to measurable effects in the polarization of 
electroluminescence in p-n junctions. \cite{zhang_2014} The electron-hole symmetry 
breaking has been confirmed by magnetoluminescence experiments. \cite{li_2014,macneill_2015} Lastly, the bands quickly display nonparabolic
dependence away from the valley minimum \cite{kormanyos_2013} which among other quantities directly affects the BC and OMM. \cite{chen_2019}

Another prevailing band structure choice is the tight-binding model for which a number of parametrizations
exist for monolayer TMDs. \cite{bromley_1972,cappelluti_2013,liu_2013,rostami_2015,fang_2015,fang_2018} 
Compared to $k\cdot p$ their agreement with first-principles 
data is over much wider range of the Brillouin zone, which comes at a price of some added formulation complexity and larger number of fitting parameters. 
Among these, arguably the simplest to use is the one by Liu {\em et al.} which is unfortunately only available for unstrained TMDs. \cite{liu_2013}
It should be noted that both $k\cdot p$ and tight-binding models warrant analytically tractable transparent physics.
In the literature there is also a vast amount of density functional theory (DFT) based results \cite{rybkovskiy_2017, jeong_2018, zhang_2017, fang_2018} 
which are highly reliable, other than the well known underestimation of the band gap by most DFT exchange-correlation functionals. 
\cite{rasmussen_2015} This entails further techniques like many-body $GW$ approximation, which yields band gaps much closer to experiments,
albeit being computationally very demanding, and so far practically inapplicable to systems beyond a few tens of atoms in the unit cell, \cite{thygesen_2017} 
making them highly undesirable for device modeling purposes.

The aim of this work is to present simple band structure options that can quantify the changes under strain in the BC and the OMM around a 
wider portion of the $K$ valleys. For this purpose, to alleviate the drawbacks of existing strained $k\cdot p$ parametrization, such as disagreement 
with the reported electron and hole effective masses as well as the band gap values, \cite{fang_2018} we develop two-band spinless and 
four-band spinful versions taking into account up-to-date first-principles and experimental data including quantum geometrical band properties, 
as will be described below. The agreement window with the ab initio and tight-binding band structures falls in the range 70-400~meV from the $K$ 
valley edge for the TMDs targeted in this work: MoS$_2$, MoSe$_2$, WS$_2$, and WSe$_2$. Moreover, we extend the tight-binding 
approach by Liu {\em et al.} \cite{liu_2013} to uniaxial and biaxial strain conditions. Based on these tools we demonstrate a doubling of 
BC and OMM for both valence band (VB) and conduction band (CB) under about 2.5\% tensile biaxial strain. We also present a simple explanation 
of how strain modifies these quantum geometrical band properties.

\section{\label{sec:Theory}Theory}
\subsection{\label{sec:2Bkp}Two-band $k\cdot p$ Hamiltonian}
For carriers near the $K$ valley edges of monolayer TMDs, the two-band $k\cdot p$ low-energy Hamiltonian $(H_0)$ which is dominated by 
the metal atom's open $d$ shell orbitals is the starting point of many studies. \cite{xiao_2012} In the presence of strain, 
characterized by the tensor components $\varepsilon_{ij}$ such that $\{i,j\}\in \{ x,y \}$, an extra term $(H_\varepsilon)$ is introduced. \cite{fang_2018} 
These two Hamiltonians are described in the Bloch basis of 
$\left\{ \ket{\bm k ,\, d_{z^2}},\, \ket{\bm k ,\, d_{x^2-y^2}+i d_{xy}}\right\}$ by  
\begin{eqnarray}
\label{H0}
H_0 & =& \frac{f_1}{2}\sigma_z+f_{2}a(k_x\sigma_x+k_y\sigma_y) ,\\
H_\varepsilon & =& f_{4}(\varepsilon_{xx}+\varepsilon_{yy})\sigma_z+f_{5}\big[(\varepsilon_{xx}-\varepsilon_{yy})\sigma_x-2\varepsilon_{xy}\sigma_y\big],
\end{eqnarray}
where $k_i$ is the wave vector Cartesian component centered around the corresponding $K$ point, $f$'s are the fitted parameters for different 
TMD materials, $a$ is the lattice constant and $\sigma_i$'s are the Pauli matrix Cartesian components. The expressions in this subsection 
specifically apply for the $+K$ valley, while those for the $-K$ valley can be obtained by complex conjugation of the matrix entries. \cite{kormanyos_2013} 
Also, we drop the constant midgap position parameters $ f_0 $ and $ f_3 $ in Ref.~\onlinecite{fang_2018}, which need to be reinstated in the study 
of heterostructures for their proper band alignment.

To account for additional features of electron-hole asymmetry, TW, and nonparabolicity we follow Korm\'{a}nyos {\em et al.} \cite{kormanyos_2013} 
by including three more terms  
\begin{equation}
H_\mathrm{2B}(\bm k)=H_0+H_\varepsilon+H_\mathrm{asym}+H_\mathrm{TW}+H_\mathrm{cubic},
\label{H2B}
\end{equation}
where,
\begin{eqnarray}
H_\mathrm{asym}=\left(\begin{array}{cc}\beta k^2&0\\0&\alpha k^2\end{array}\right),
\label{H_as}
\end{eqnarray}
\begin{eqnarray}
H_\mathrm{TW}=\kappa\left(\begin{array}{cc}0&k_+^2\\k_-^2&0\end{array}\right),
\label{H_TW}
\end{eqnarray}
\begin{eqnarray}
H_\mathrm{cubic}=\frac{\eta}{2}k^2\left(\begin{array}{cc}0&k_-\\k_+&0\end{array}\right),
\label{Hcub}
\end{eqnarray}
and $k_{\pm}=k_x \pm ik_y$, the parameters $\alpha$ and $\beta$ describe the breaking of the electron-hole symmetry, 
whereas $\kappa$ is responsible for the TW of the isoenergy contours, and $H_\mathrm{cubic}$ serves to improve the 
fit further away from the $K$ point. \cite{kormanyos_2013}

\subsection{Three-band tight-binding Hamiltonian}
The two-band $k\cdot p$ approach is inevitably restricted to the vicinity of the $K$ points. To extend it over a wider part of 
the Brillouin zone the number of bands need to be increased considerably.\cite{rybkovskiy_2017} For the sake of simplicity, we rather 
prefer the three-band tight-binding (TB) approach which provides a full-zone band structure fitted to the first-principles 
data, where in the case of up to third nearest neighbor interactions 
19 fitting parameters are involved. \cite{liu_2013} It assumes the Bloch basis of 
$\left\{\ket{\bm k ,\, d_{z^2}}\, , \ket{\bm k ,\,d_{xy}}\, , \ket{\bm k ,\,d_{x^2-y^2}}\right\}$ 
coming from the atomic orbitals that largely contribute to the VB and CB edges of TMDs. \cite{kormanyos_2013}
The matrix representation of the Hamiltonian takes the form
\begin{equation}
H_{0}=\left(\begin{array}{ccc} V_{0} & V_{1} & V_{2} \\V_1^* & V_{11} & V_{12} \\V_2^* & V_{12}^* & V_{22} \end{array}\right),
\end{equation}
where $V_0, \, V_1, \, V_2, \, V_{11}, \, V_{12}, \, V_{22}$ are the TB matrix elements; for their detailed expressions 
we refer to Ref.~\onlinecite{liu_2013}.
Though this Hamiltonian is highly satisfactory it is for unstrained TMDs. We remedy this by 
the two-band deformation potentials proposed by Fang {\em et al.} \cite{fang_2018} 
that we also use in our $k\cdot p$ theory in Sec.~\ref{sec:2Bkp}. So, the strain is embodied into the three-band TB Hamiltonian as
\begin{equation}
H_\varepsilon=\left(\begin{array}{ccc} e_a & e_b & e_b \\e_b & -e_a & 0 \\e_b & 0 & -e_a \end{array}\right),
\label{H_TBstr}
\end{equation}
where
\begin{eqnarray}
e_a & = & f_4 \left( \varepsilon_{xx}+\varepsilon_{yy}\right), \\
e_b & = & f_5 \left( \varepsilon_{xx}-\varepsilon_{yy}\right) .
\end{eqnarray}
Here, our simplistic approach lends itself to a number of restrictions.
Even though this TB is a three-band model, the deformation potentials are only available for the two-band case (highest VB and 
the lowest CB). \cite{fang_2018} Therefore, we expand it to the two-dimensional subspace formed by $\ket{\bm k ,\,d_{xy}}$ and 
$\ket{\bm k ,\,d_{x^2-y^2}}$ which define the highest VB and the first-excited CB around the $K$ valleys, 
while neglecting the strain coupling between them. Its form (Eq.~(\ref{H_TBstr})) complies with the TB $d-d$ sector 
deformation coupling of monolayer TMDs. \cite{pearce_2016} As another remark, here strain only acts through the uniaxial and 
biaxial components, with no involvement of the shear strain ($\varepsilon_{xy}=\varepsilon_{yx}$). 
In fact, it has been shown for this level of theory that 
the latter is only responsible for a rigid shift of the band extrema. \cite{fang_2018,aas_2018}

\begin{figure}
\includegraphics[width=0.52\textwidth]{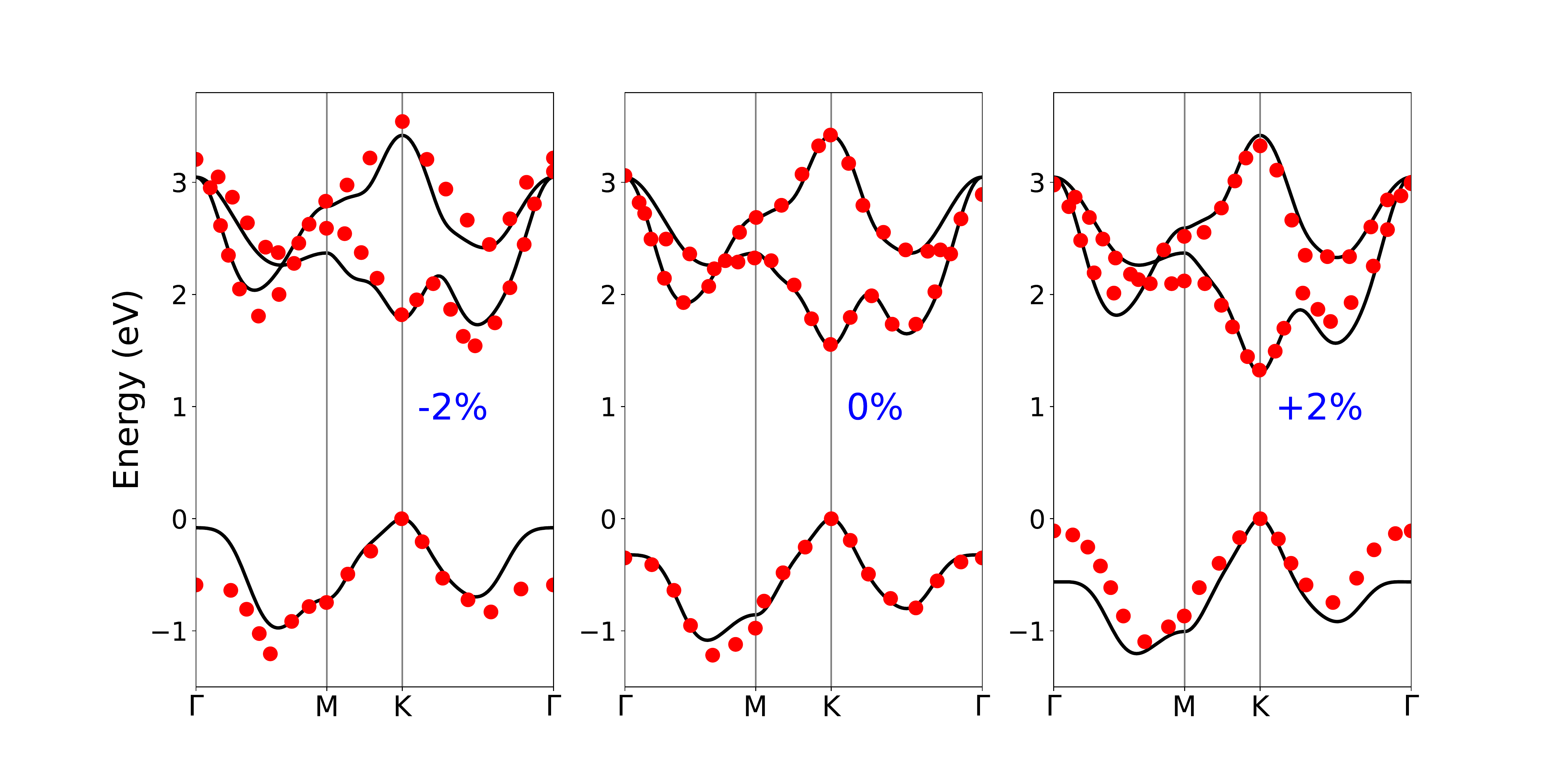}
\caption{\label{fig1} Comparison of first-principles band structure (red dots) \cite{fang_2018} with the TB results (black line) 
for WSe$_2$ under unstrained (0\%) and $\pm$~2\% biaxial strain cases. Energy reference is set to VB maximum for each case.} 
\end{figure}

To test the validity of this simple strain extension, in Fig.~\ref{fig1} we compare it with the first-principles 
band structure results for WSe$_2$ under $\pm$~2\% biaxial, and unstrained cases. \cite{fang_2018} As intended, the agreement
around the $K$ valley is quite satisfactory, whereas disagreement sets in away from this region especially toward the $\Gamma$ point.
Apparently, $\Gamma$ and $K$ valleys have different signs for the deformation potentials causing a direct to indirect transition 
under compressive strain. Thus, it cannot be represented with only that of a single (i.e., $K$) valley. As a matter of fact, 
even for the unstrained case the original TB fitting has deficiencies around the $\Gamma$ point. \cite{liu_2013} 
These limitations will not be of practical concern for this work as the geometrical band properties that we are interested in are 
localized around the $K$ point, and vanish toward the $\Gamma$ point due to symmetry considerations. \cite{feng_2012}

\subsection{Berry curvature and orbital magnetic moment}
In the absence of an external magnetic field TMDs respect the time reversal symmetry, but inversion symmetry is broken in monolayers or odd 
number of layers as has been independently confirmed by recent experiments. \cite{mak_2014,wu_2019}
Therefore, in monolayer TMDs BC has a non-zero value with opposite sign in $K$ and $-K$ valleys connected by a time-reversal operation. \cite{xiao_2010}
For a chosen band with label $n$ it can be calculated without reference to other bands using
\begin{equation}
\Omega_{n,z}(\bm k)=-2\,\mathrm{Im}\left\langle\partial_{k_x} u_{n{\bm k}} | \partial_{k_y} u_{n{\bm k}}\right\rangle,
\end{equation}
where $z$ is the direction perpendicular to monolayer plane, $\ket{u_{n{\bm k}}}$ is the cell-periodic part of the Bloch function 
at wave vector ${\bm k}$. 
\cite{vanderbilt_2018} Another geometric band property is the OMM which is also a pseudovector given by
\begin{equation}
\mu_{n,z}(\bm k)=2\frac{\mu_B m_0}{\hbar^2}\mathrm{Im}\bra{\partial_{k_x} u_{n{\bm k}}} \left[ H(\bm k)-E_n(\bm k) \right] \ket{\partial_{k_y} u_{n{\bm k}}},
\end{equation} 
where $\mu_B$ is the Bohr magneton, $m_0$ is the free-electron mass, and $E_n(\bm k)$ is the energy of the band $n$ at the wave vector $\bm k$.
\subsection{\label{sec:fit}Fitting Procedure and Data References}
%
\begin{figure}
\subfloat[][]{
\includegraphics[width=0.5\textwidth]{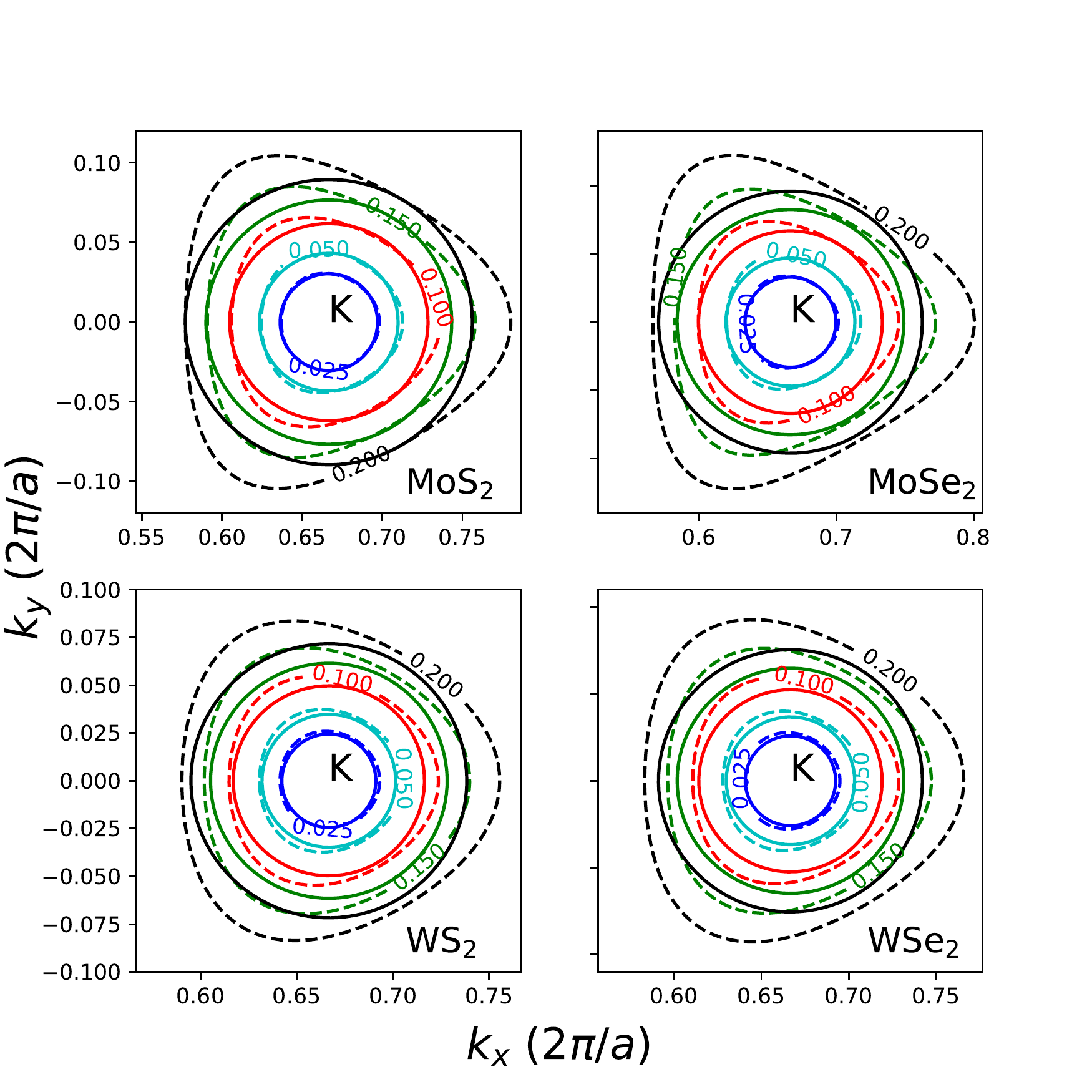}
\label{fig2a}}
\par
\subfloat[][]{
\includegraphics[width=0.5\textwidth]{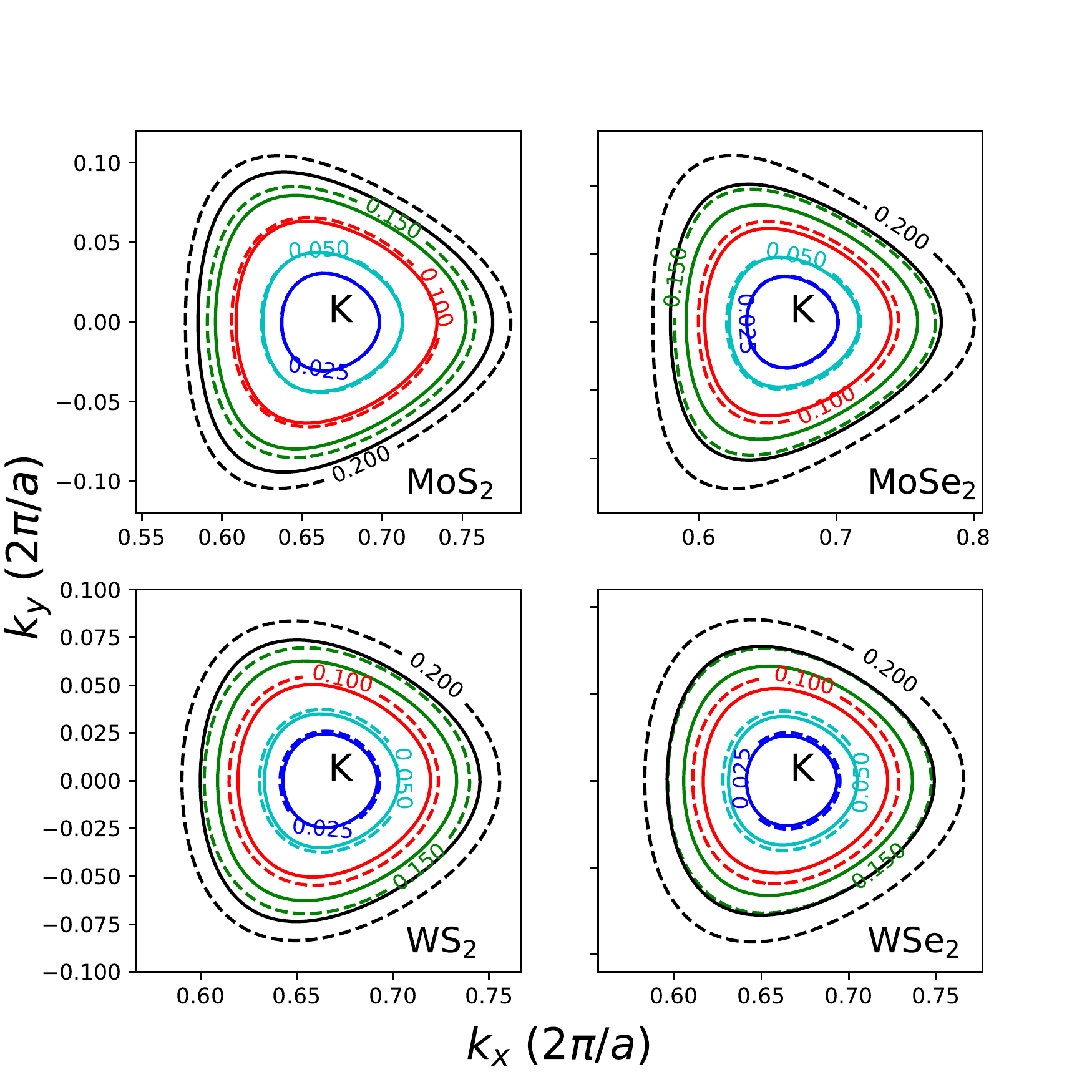}
\label{fig2b}}
\caption{\label{fig2} Isoenergy contours of the VB (at energies indicated in eV) for different TMDs 
from TB (dashed) and $k\cdot p$ (solid) models, where the latter excludes (a), or includes (b) the TW effect.
For these plots, $\Gamma$ point has been taken as the origin for $\bm k$. } 
\end{figure}
Our two-band $k\cdot p$ model depends on the following parameters:
$ a $, $ f_1 $, $ f_2 $, $ f_4 $, $ f_5 $, $\alpha$, $ \beta $, $ \kappa $, $ \eta $. The lattice constant, $a$ is taken from 
DFT (GGA) model calculations (Table~\ref{tableI}).\cite{liu_2013}  For the remaining eight parameters, rather than going through a formidable 
simultaneous optimization in such a high-dimensional parameter space, a sequential fitting is possible as follows.
$ f_1 $ determines the free-particle band gap which we fit to the corresponding experimental $E_g$ using scanning tunneling spectroscopy
data (i.e., without the excitonic contributions) listed in the recent review (Table~\ref{tableI}). \cite{wang_2018} 
For $f_2$, we make use of the fact that the BC expression at the $K$ point 
simplifies to $\Omega(K)=\pm 2 (f_2 a/f_1)^2$. \cite{xiao_2007} We fit the average of this quantity for the lowest spin-allowed 
transitions in $K$ valley ($|\bar{\Omega}|=(|\Omega_\mathrm{CB}(K)|+|\Omega_\mathrm{VB}(K)|)/2$) to the first-principles results (Table~\ref{tableI})
\cite{feng_2012} which resolves the $ f_2 $ parameter. $ f_4 $ and $ f_5 $ characterize 
the strain and they are directly acquired from Ref.~\onlinecite{fang_2018} without any change. After these set of parameters for $H_0$, 
we move to $H_\mathrm{asym}$ for $\alpha$ and $ \beta $. We readily extract these from the reported effective masses (Table~\ref{tableI}). 
\cite{kormanyos_2015} As a two-band model, again we select the effective masses of lowest spin-allowed VB-CB transitions in the fitting procedure. 

\begin{table}[H]
\caption{\label{tableI}Target data for $k\cdot p$ parameters. 
Lattice constant ($a$),\cite{liu_2013} single-particle band gap ($E_g$),\cite{wang_2018} average of BC ($\bar{\Omega}$),\cite{feng_2012} 
VB ($m^*_\mathrm{vb}$) and CB ($m^*_\mathrm{cb}$) effective masses.\cite{kormanyos_2015} 
$E_g$, and the two-band $k\cdot p$ parameters are based on spin-allowed lowest energy VB-CB transition between spin-down ($\downarrow$) 
states according to Ref.~\onlinecite{kormanyos_2015}.}
\begin{ruledtabular}
 \begin{tabular}{lcccc}
     Materials & MoS$_2$ & MoSe$_2$  & WS$_2$ & WSe$_2$\\ 
         \hline
     $a$ (\AA) & 3.190 & 3.326 & 3.191 & 3.325\\
     $E_g$ (eV) & 2.15 & 2.18 & 2.38 & 2.20\\
     $|\bar{\Omega}|$ (\AA$^2$) & 10.43 & 10.71 & 16.03 & 17.29\\
     $m^*_\mathrm{vb,\downarrow}$ $(m_0)$ & -0.54 & -0.59 & -0.35 & -0.36\\
     $m^*_\mathrm{cb,\downarrow}$ $(m_0)$ & 0.43 & 0.49 & 0.26 & 0.28\\
     $m^*_\mathrm{vb,\uparrow}$ $(m_0)$ & -0.61 & -0.7 & -0.49 & -0.54\\
     $m^*_\mathrm{cb,\uparrow}$ $(m_0)$ & 0.46 & 0.56 & 0.35 & 0.39\\
\end{tabular}
\end{ruledtabular}
\end{table}

Figure~\ref{fig2} compares the isoenergy contours plotted using Eqs.~(\ref{H0}) and (\ref{H_as}) (solid lines), with the TB model 
calculations \cite{liu_2013} (dashed lines). Each color corresponds to a different amount of excess energy as measured from the VB 
$K$ valley edge (i.e. VB maximum). Figure~\ref{fig2}~(a) displays the case without
TW in $k\cdot p$ calculations resulting in circular curves. By adding Eq.~(\ref{H_TW}) to the previous Hamiltonian 
(Eqs.~(\ref{H0}) and (\ref{H_as})) TW effect on the isoenergy contours emerges (Fig.~\ref{fig2}~(b)). We fix the $\kappa$ 
parameter by fitting the $k\cdot p$ to the TB model at the $100$~meV isoenergy contour. 
Finally to extract the $\eta$ parameter we fit the band structure of different TMDs calculated from Eq.~(\ref{H2B}) to the recent 
DFT data. Our final two-band $k\cdot p$ parameter set for the four TMDs is presented in Table~\ref{tableII}.

\begin{figure}
\includegraphics[width=0.5\textwidth]{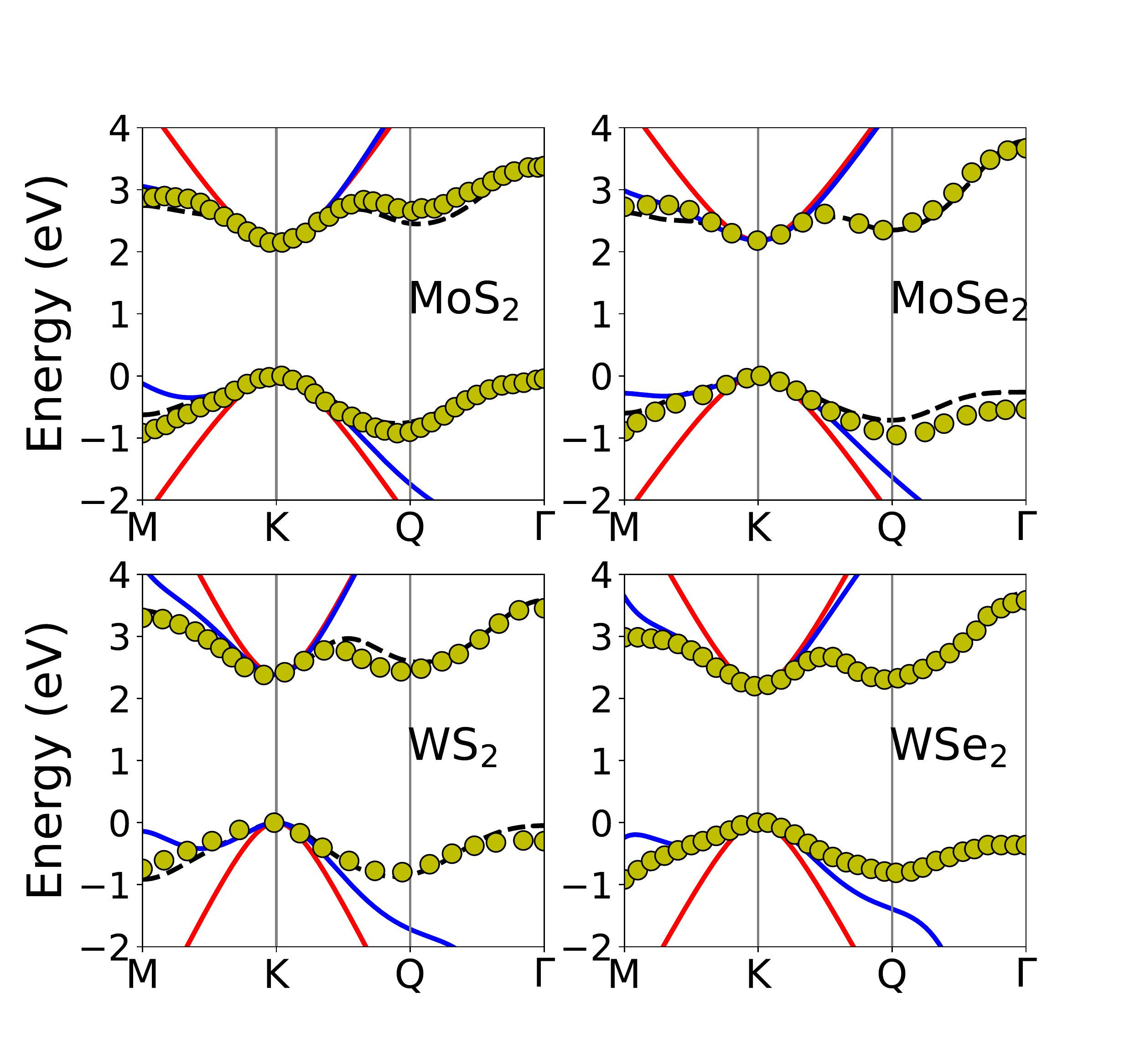}
\caption{\label{fig3} $k\cdot p$ band structure of monolayer TMDs with (blue),  and without (red) taking into account the electron-hole asymmetry, 
TW and nonparabolic effects, compared with the TB calculations (black dashed lines), and DFT results (yellow dots) collected from various references. 
\cite{rybkovskiy_2017, jeong_2018, zhang_2017, fang_2018} To facilitate the comparison, VB maxima are set to zero energy, and the band gaps in each 
case is corrected to the values in Table~\ref{tableI}.}
\end{figure}

Figure~\ref{fig3} contrasts band structure of different TMDs from Eq.~(\ref{H0}) (red curves), and including additional terms in 
Eq.~(\ref{H2B}) using our $k\cdot p$ fitted parameters (blue curves) along with the DFT values (yellow dots). 
\cite{rybkovskiy_2017, jeong_2018, zhang_2017, fang_2018} Furthermore, we plot TB band structures \cite{liu_2013} (black dashed curves) in this 
figure to assess how precise is our two-band model. Notably, DFT and TB model are in excellent agreement. Also, the blue curves 
from the two-band $k\cdot p$ Hamiltonian, $H_\mathrm{2B}$ calculations approach to DFT and TB model results around $K$ valley which assure 
the benefit of these additional terms (Eqs.~(\ref{H_as})$-$(\ref{Hcub})) in Eq.~(\ref{H0}). The energy range within $10$~meV 
agreement with TB and DFT  data \cite{rybkovskiy_2017, jeong_2018, zhang_2017, fang_2018} are included in Table~\ref{tableII}. 
The narrowest among these is for WS$_2$ CB which is $70$~meV, and widest for MoSe$_2$ for VB with 400~meV, both as measured 
from the respective band edges. Thus, for intravalley transport these can suffice, except for the hot carrier regime for which 
we advise to switch to TB model.
\begin{table}[H]
\caption{\label{tableII}Fitted two-band $k\cdot p$ parameters, and within-10~meV-agreement window for VB and CB.}
\begin{ruledtabular}
 \begin{tabular}{l c c c c}
   TMD & MoS$_2$ & MoSe$_2$  & WS$_2$ & WSe$_2$\\ 
       \hline
       $f_1$ (eV) & 2.15 & 2.18 &2.38 & 2.2\\
       $f_2$ (eV) & 1.54 &  1.52 & 2.11 & 1.95\\
       $f_4$ (eV) & -2.59 & -2.28 & -3.59 & -3.02\\
       $f_5$ (eV) & 2.2 & 1.84 & 2.27 & 2.03\\
       $\alpha$ (eV$\cdot$\AA$^2$) & 4.16 & 5.22 & 8.2 & 8.43\\
       $\beta$ (eV$\cdot$\AA$^2$) & -2.35 & -3.9 & -4.43 & -5.4\\
       $\kappa$ (eV$\cdot$\AA$^2$) & -1.9 & -1.8 & -2.2 & -2\\
       $\eta$ (eV$\cdot$\AA$^3$) & 6 & 8 & 14 & 18\\
       VB Fit Range (meV) & 350 & 400 & 200 & 100\\ 
       CB Fit Range (meV) & 115 & 170 & 70 & 90\\
 \end{tabular}
 \end{ruledtabular}
\end{table}

\subsection{Spin-dependent four-band $k\cdot p$ Hamiltonian}
Due to the presence of heavy metal atoms in TMDs, the spin-orbit interaction is quite strong \cite{pearce_2016} in contrast to 
for instance, monolayer graphene and hBN.\cite{fang_2018} The spin-dependent effects in TMDs are commonly incorporated within the 
spin-diagonal and wave vector-independent approximation.\cite{xiao_2012,liu_2013,kormanyos_2013,kormanyos_2015}  
Thus, we first generalize the two-band Hamiltonian of Eq.~(\ref{H2B}) into a form with spin-dependent diagonal entries as
\begin{equation}
H_\mathrm{2B}=\left(\begin{array}{cc}h_{11} & h_{12}\\h^*_{12} & h_{22}\end{array}\right)
\longrightarrow ~~
H^{\downarrow,\uparrow}_\mathrm{2B}=\left(\begin{array}{cc}h^{\downarrow,\uparrow}_{11} & h_{12}\\h^*_{12} & h^{\downarrow,\uparrow}_{22}\end{array}\right),
\label{H2Bspin}
\end{equation}
by modifying only the electron-hole asymmetry contribution in Eq.~(\ref{H_as}) so that it becomes
\begin{eqnarray}
H^{\downarrow,\uparrow}_\mathrm{asym}=k^2\left(\begin{array}{cc}\beta^{\downarrow,\uparrow} &0\\0&\alpha^{\downarrow,\uparrow} \end{array}\right),
\label{H_asspin}
\end{eqnarray}
where the fitted $\beta^{\downarrow,\uparrow}$ and $\alpha^{\downarrow,\uparrow}$ to the corresponding effective mass values are tabulated in 
Table~\ref{tableIII}. We keep the remaining two-band $k\cdot p$ parameters ($f_1, f_2, f_4, f_5, \kappa, \eta$) as in Table~\ref{tableII}, 
and in this way we do not inflate the number of fitting parameters significantly.

With these ingredients the two-band formalism is extended into both spin channels that results in the four-band $k\cdot p$ Hamiltonian 
which is expressed in the Bloch basis ordering of $\left\{ \ket{\bm k ,\, d_{z^2},\uparrow},\, 
\ket{\bm k ,\, d_{z^2},\downarrow},\, \ket{\bm k ,\, d_{x^2-y^2}+i d_{xy},\uparrow},\, \ket{\bm k ,\, d_{x^2-y^2}+i d_{xy},\downarrow} \right\}$ as
\begin{equation}
H_\mathrm{4B} = \left(\begin{array}{cccc} h^\uparrow_{11} & 0 & h_{12} & 0\\ 0 & h^\downarrow_{11} & 0 & h_{12} \\
h^*_{12} & 0 & h^\uparrow_{22} & 0 \\ 0 & h^*_{12} & 0 & h^\downarrow_{22} \end{array}\right)
+ \tau \left(\begin{array}{cccc} -\Delta^\mathrm{so}_\mathrm{cb} & 0 & 0 & 0\\ 0 & 0 & 0 & 0\\ 0 & 0 & -\Delta^\mathrm{so}_\mathrm{vb} & 0\\
0 & 0 & 0 & 0 \end{array}\right),
\label{H4B}
\end{equation}
where $\tau$ is the valley index with value $+1$ ($-1$) for the $+K$ ($-K$) valley, and $\Delta^\mathrm{so}_\mathrm{cb}$ ($\Delta^\mathrm{so}_\mathrm{vb}$)
is the CB (VB) spin splitting as listed in Table~\ref{tableIII}. 

\begin{table}[H]
\caption{\label{tableIII} Additional spin-dependent $k\cdot p$ parameters required for the four-band Hamiltonian. CB and VB spin splittings are taken from
spin-polarized DFT band structure.\cite{kormanyos_2015} $\alpha^\downarrow$ and $\beta^\downarrow$ values coincide with the two-band values in Table~\ref{tableII}.}
\begin{ruledtabular}
 \begin{tabular}{l c c c c}
   TMD & MoS$_2$ & MoSe$_2$  & WS$_2$ & WSe$_2$\\ 
       \hline
       $\Delta^\mathrm{so}_\mathrm{cb}$ (meV) & -3 & -22 & 32 & 37\\
       $\Delta^\mathrm{so}_\mathrm{vb}$ (meV) & 148 &  186 & 429 & 466\\
       $\alpha^\downarrow$ (eV$\cdot$\AA$^2$) & 4.16 & 5.22 & 8.2 & 8.43\\
       $\beta^\downarrow$ (eV$\cdot$\AA$^2$) & -2.35 & -3.9 & -4.43 & -5.4\\
       $\alpha^\uparrow$ (eV$\cdot$\AA$^2$) & 4.23 & 5.22 & 8.58 & 8.85\\
       $\beta^\uparrow$ (eV$\cdot$\AA$^2$) & -2.2 & -3.86 & -5.47 & -6.15\\
 \end{tabular}
 \end{ruledtabular}
\end{table}

\begin{figure}
\includegraphics[width=0.5\textwidth]{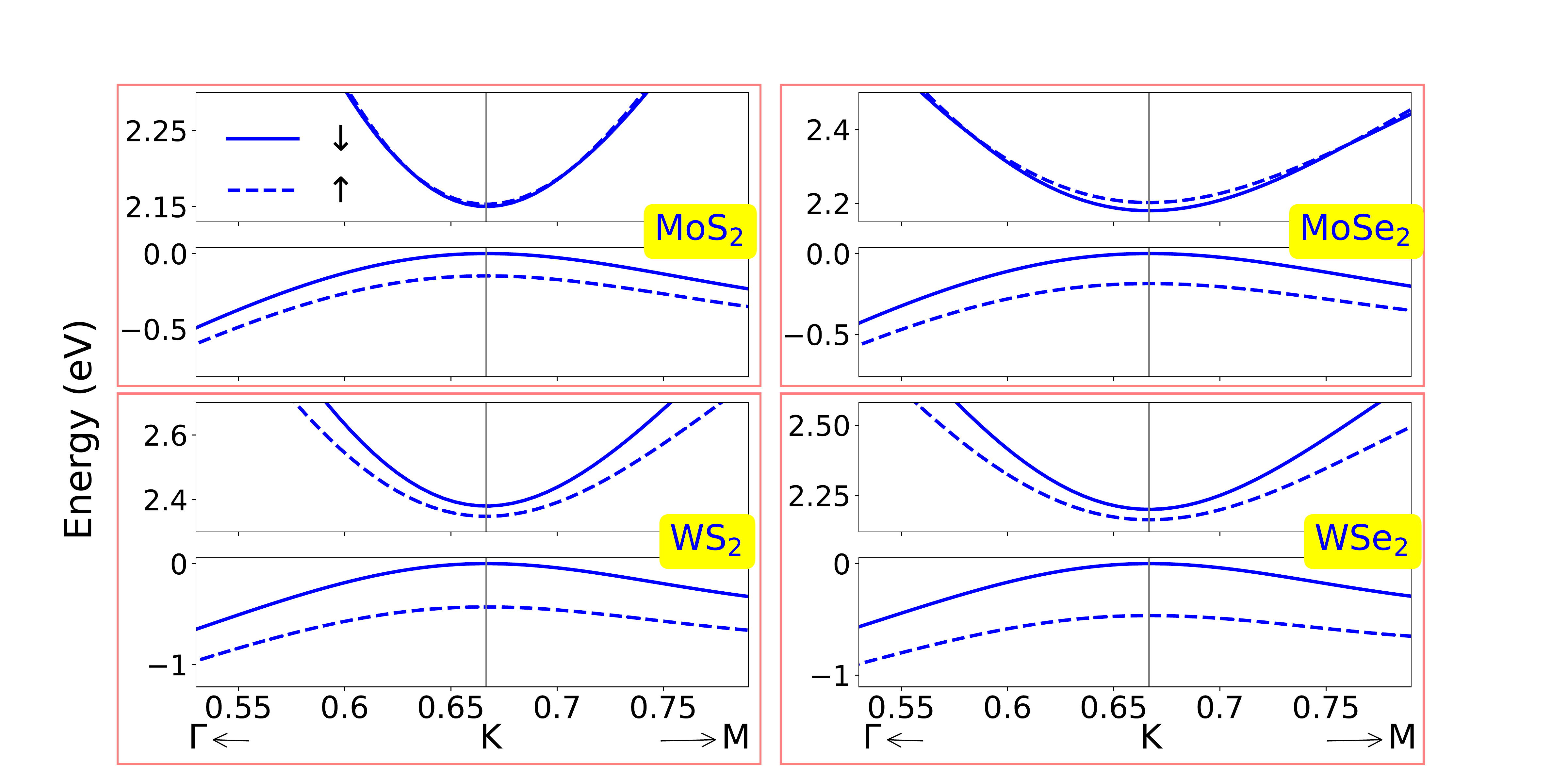}
\caption{\label{fig4} Four-band $k\cdot p$ band structure of monolayer TMDs around the $K$ valley for 
spin $\downarrow$ (solid) and spin $\uparrow$ (dashed) bands. Abscissae are in units of $2\pi/a$. 
Spin-dependent parameters are listed in Table~\ref{tableIII}, and the remaining spin-independent parameters 
($f_1, f_2, f_4, f_5, \kappa, \eta$) are used from Table~\ref{tableII}.}
\end{figure}

Figure~\ref{fig4} shows the spin-dependent band structure of the monolayer TMDs around the $K$ valley.
As the spin-dependent parameters in Eq.~(\ref{H4B}) reside on the diagonal entries, in this level of approximation
spin remains to be a good quantum label. \cite{xiao_2012} Another convenience of this approach is that
the aforementioned two-band model directly corresponds to the spin-$\downarrow$ sector of the four-band Hamiltonian. Therefore,
when lowest-lying spin-allowed transitions (as in the so-called $A$-excitons) are of interest,\cite{wang_2018} 
the spinless two-band variant in Sec.~\ref{sec:2Bkp} can be employed.

\section{\label{sec:Result}Results and Discussion}
\begin{figure}
\subfloat[][]{
\includegraphics[width=0.5\textwidth]{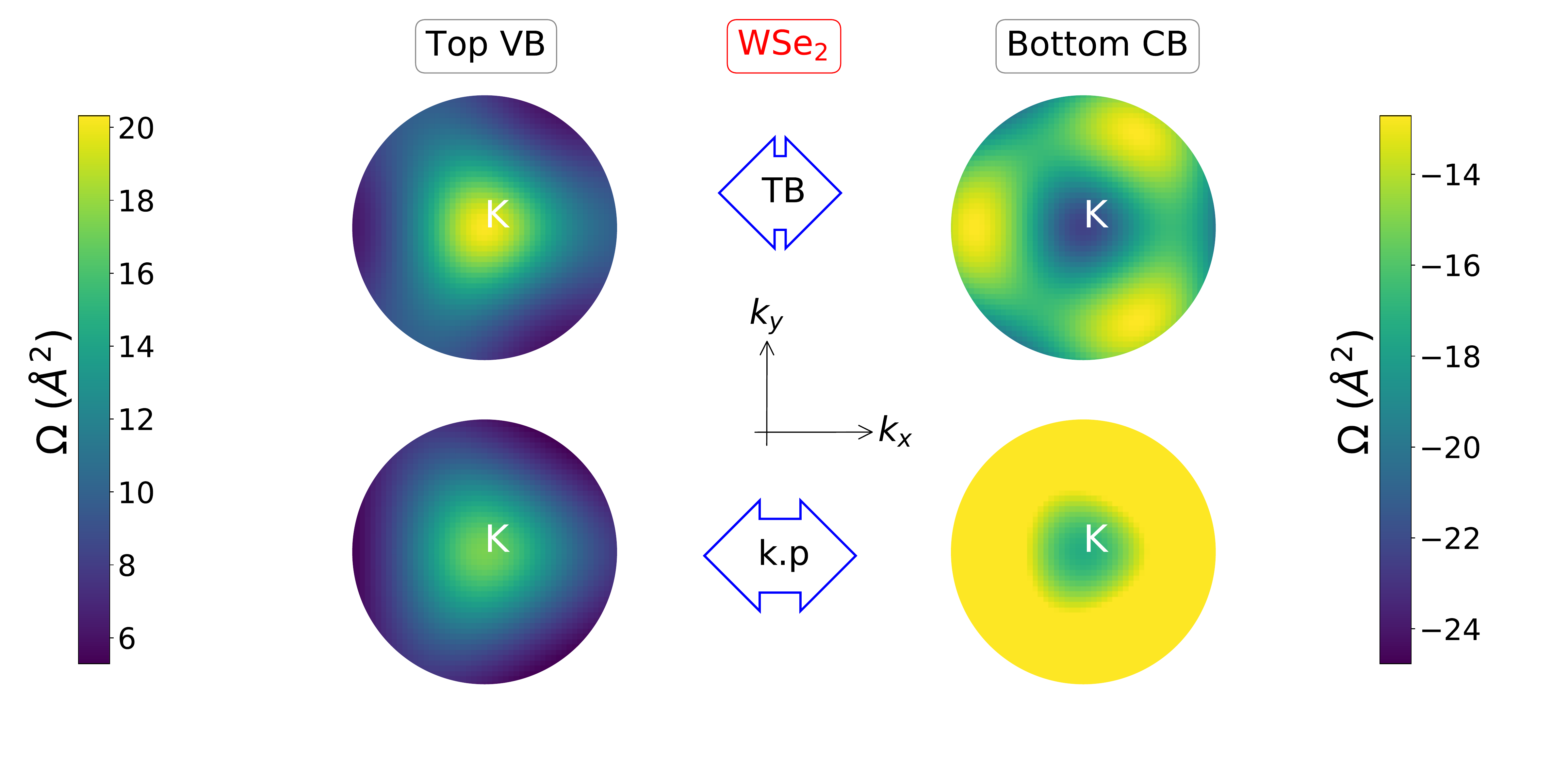}
\label{fig5a}}
\par
\subfloat[][]{
\includegraphics[width=0.5\textwidth]{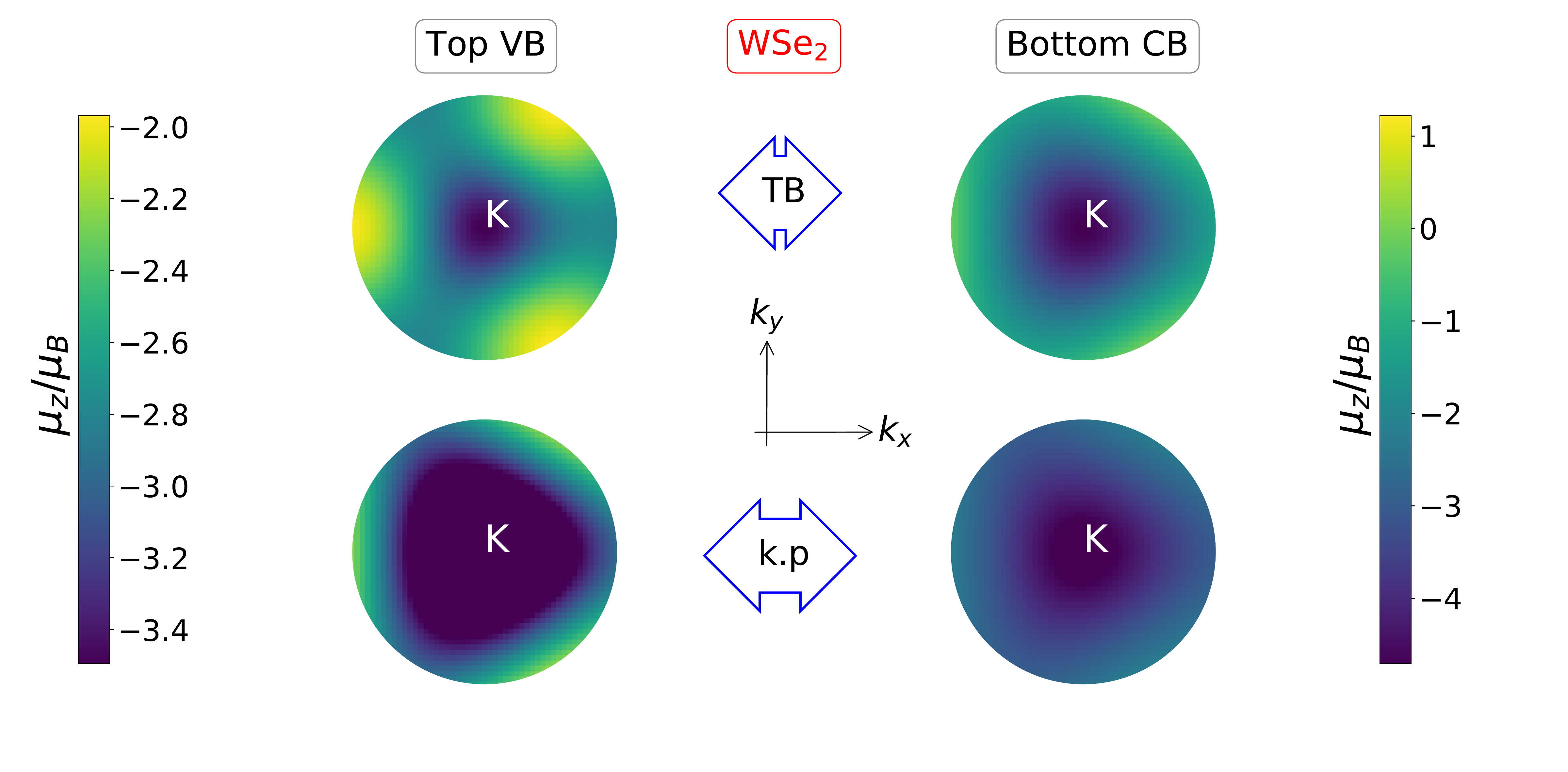}
\label{fig5b}}
\caption{\label{fig5}(a) BC and (b) OMM of (spin $\downarrow$) CB and VB for an unstrained monolayer 
WSe$_2$. The part of the Brillouin zone centered at the $K$ point over a radius of $0.12\times 2\pi/a$ is shown,
maintaining the same axial orientation as in Fig.~\ref{fig2}.} 
\end{figure}

To demonstrate several aspects of the geometric band properties we choose monolayer WSe$_2$ as the prototypical TMD material, and
focus on the bands with the narrowest spin-allowed bandgap transition which corresponds to spin--$\downarrow$ sector at the $K$ 
valley (solid lines in Fig.~\ref{fig4}) which essentially reduces the computational task to 
the two-band $k\cdot p$, and the three-band TB cases, as mentioned above. Starting with the unstrained case in Fig.~\ref{fig5}, 
the top VB and bottom CB behaviors for both of these models are in qualitative agreement around $K$ valley edge,
with the variation in the TB being wider for both geometric quantities. The significance of TW on these can be clearly 
observed together with the fact that BC toggles sign between VB and CB while this is not the case for the OMM. 
\subsection{\label{sec:strain effect}Effects of strain}
\begin{figure}
\subfloat[][]{
\includegraphics[width=0.5\textwidth]{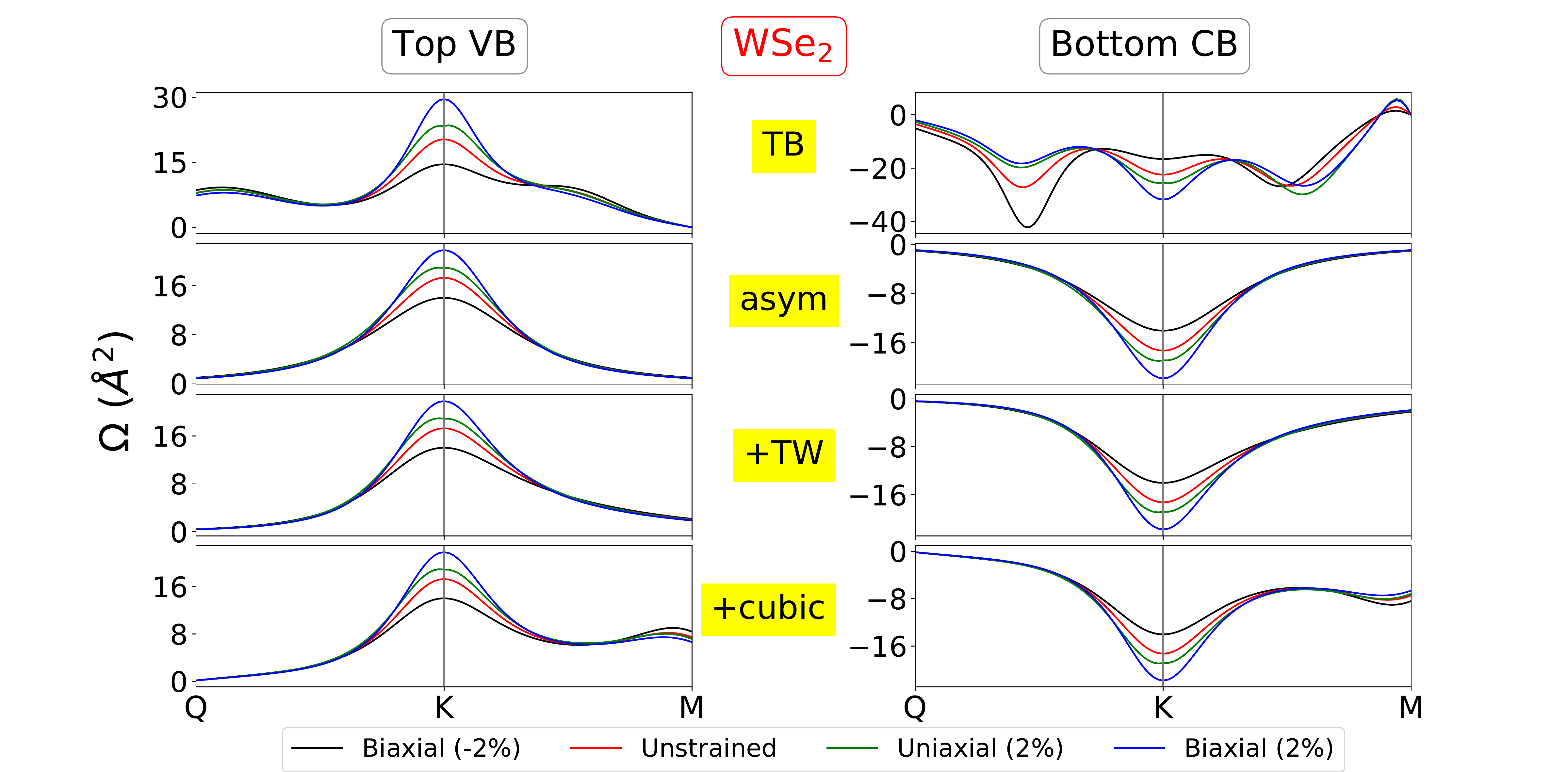}
\label{fig6a}}
\par
\subfloat[][]{
\includegraphics[width=0.5\textwidth]{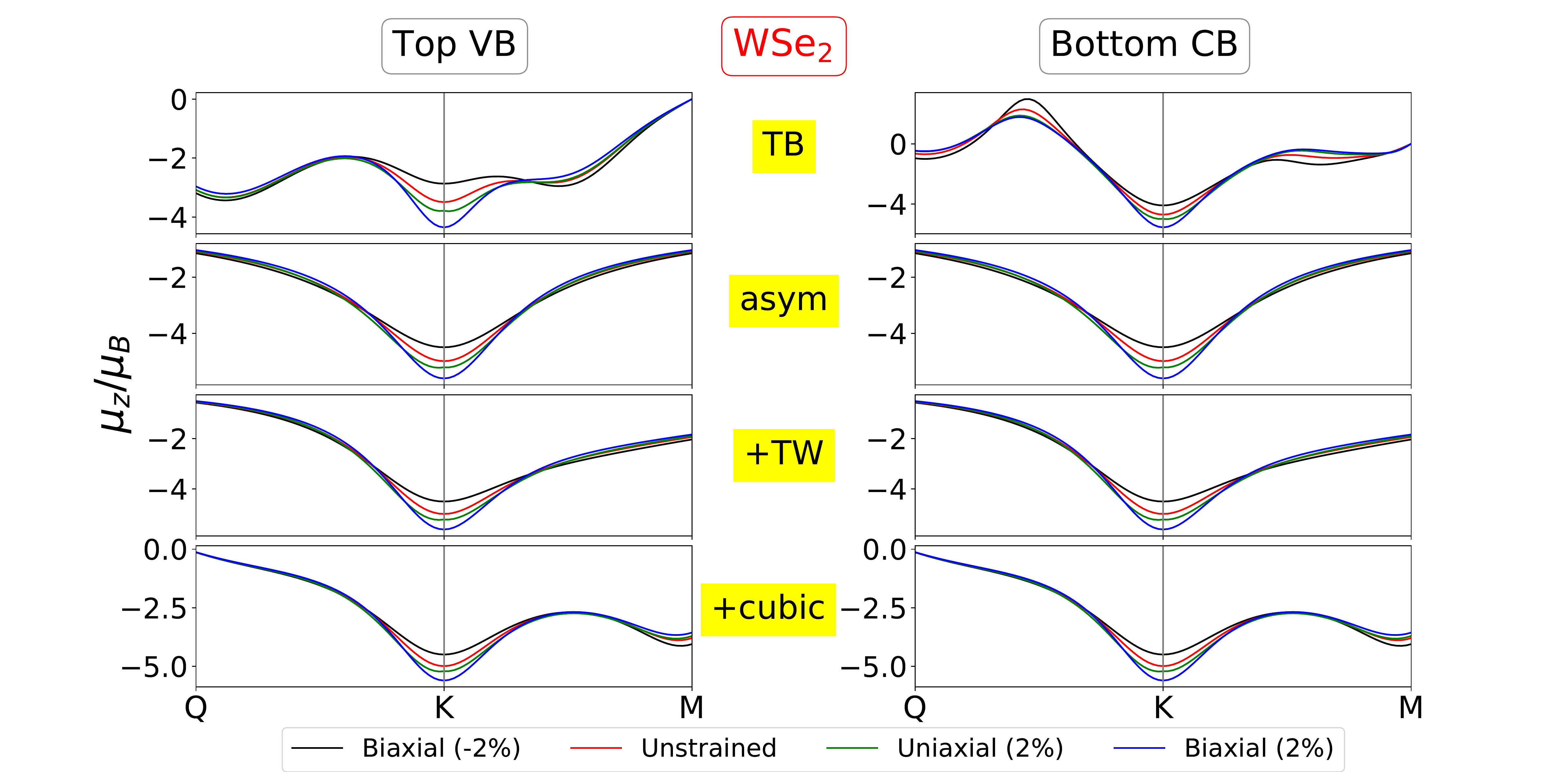}
\label{fig6b}}
\caption{\label{fig6}Effect of strain for a monolayer WSe$_2$ on (a) BC and (b) OMM of (spin $\downarrow$) CB and VB according 
to TB (top rows) and $k\cdot p$ (remaining rows) models, where for the latter the effect of each additional 
term (Eqs.~(\ref{H_as})$-$(\ref{Hcub})) is shown.} 
\end{figure}

Figure~\ref{fig6} shows the effects of strain on the (a) BC, and (b) OMM for the monolayer WSe$_2$ over the $Q-K-M$ path 
within the Brillouin zone, where the $Q$ point lies exactly at midway between the $\Gamma$ and $K$ points.
First considering the TB results, the geometric properties are seen to be inflated as the strain changes from compressive 
to tensile nature. However, this simple behavior is localized to the $K$ valley, especially for the VB. 
In the case of the CB, the variation gets reversed beyond the halfway between the $Q-K$ panel, due to the satellite CB valley 
at the $Q$ point.\cite{kormanyos_2015} Switching to $k\cdot p$ results, in the vicinity of $K$ valley they display a 
behavior close to TB but again with somewhat reduced amplitudes. 
The incremental contribution of each term in the Hamiltonian (Eqs.~(\ref{H_as})$-$(\ref{Hcub})) indicates that the cubic term 
actually deteriorates the agreement with TB toward the $M$ point by introducing an extra curvature for both VB and CB, 
yet it was observed in Fig.~\ref{fig3} to have a positive impact on the band structure for the same point.
\begin{figure}[!ht]
\subfloat[][]{
\includegraphics[width=0.50\textwidth]{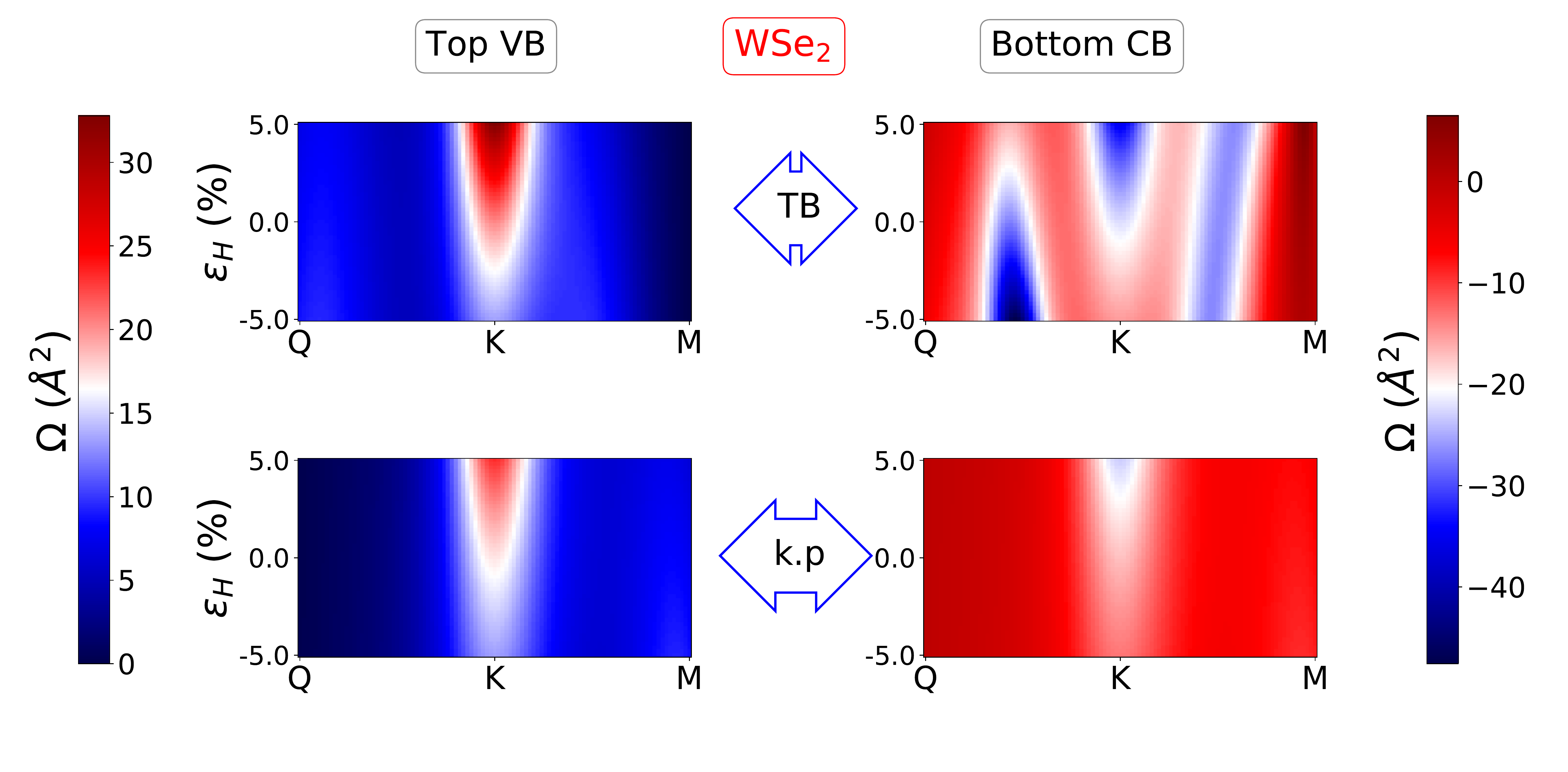}
\label{fig7a}}
\par
\subfloat[][]{
\includegraphics[width=0.5\textwidth]{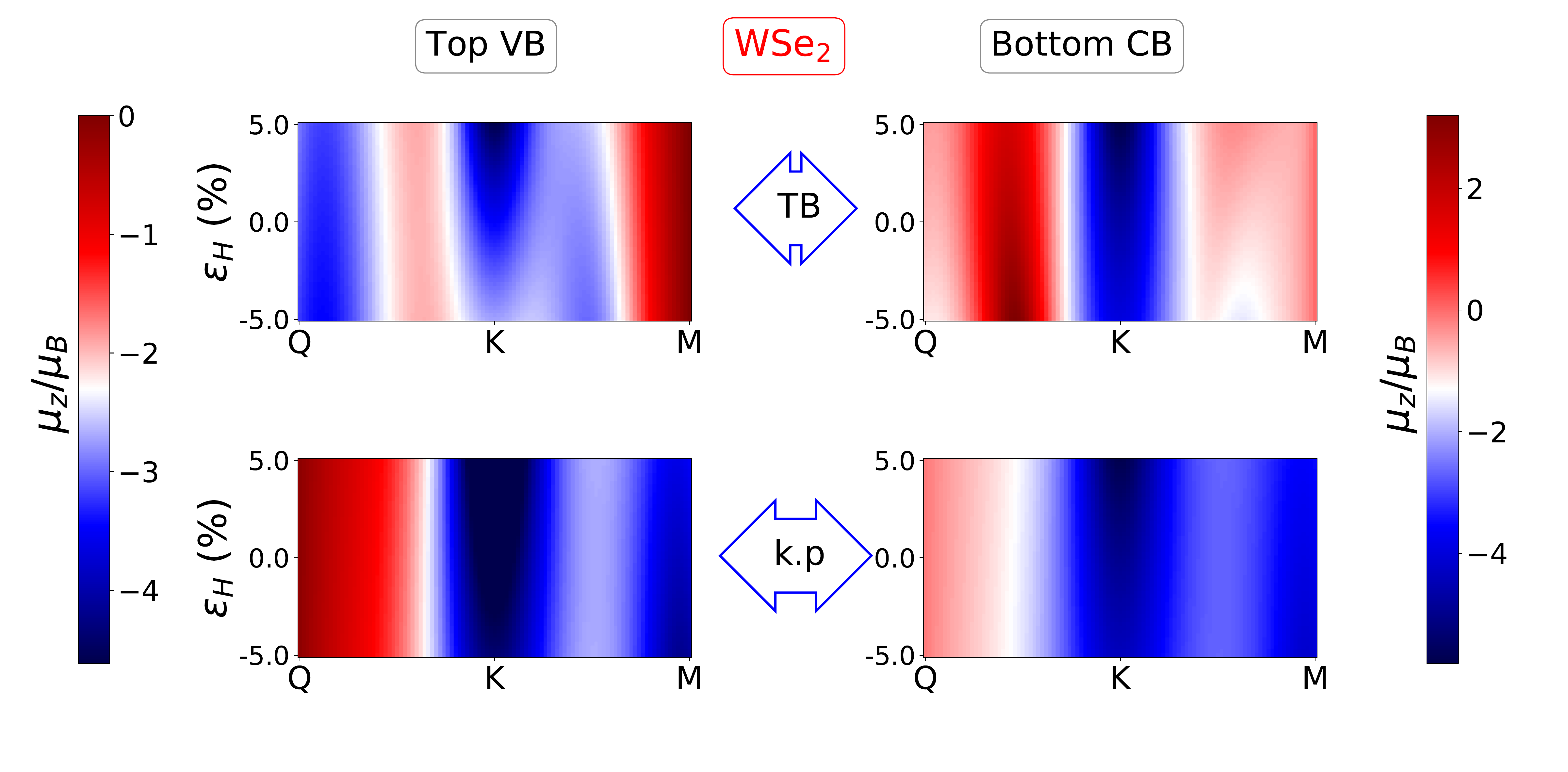}
\label{fig7b}}
\caption{\label{fig7}Effect of (compressive/tensile) hydrostatic strain, $\varepsilon_H=\varepsilon_{xx}+\varepsilon_{yy}$
 on (a) BC and (b) OMM of (spin $\downarrow$) CB and VB for a monolayer WSe$_2$.} 
\end{figure}

These traits are more clearly demonstrated in Fig.~\ref{fig7} where the continuous tunability of both BC and OMM under hydrostatic strain
$\varepsilon_H=\varepsilon_{xx}+\varepsilon_{yy}$ is displayed. Once again, $k\cdot p$ while in qualitative agreement with TB around the $K$ 
valley, it cannot reproduce the broad variations; particularly for the CB the $Q-K$ panel is not satisfactory. 
As a matter of fact a separate $k\cdot p$ Hamiltonian needs to be invoked to replicate the correct behavior around the CB $Q$ 
valley.\cite{kormanyos_2015} Apart from these discrepancies at regions 
with relatively low curvature, both techniques reveal that the $K$ point geometric band properties can be doubled with respect 
to unstrained values by about +5\% hydrostatic strain. 

We can offer a simple explanation for these increased geometrical band properties under tensile hydrostatic strain by
making use of two-band electron-hole symmetric analytical expressions \cite{xiao_2007,chen_2019} for the $K$ point: 
$\Omega_z=\pm 2(f_2 a/E_g)^2$, $\mu_z=\mu_B m_0/m^*$,
where the strained band gap \cite{aas_2018} $E_g=f_1+2f_4\varepsilon_H$, and the strained effective mass \cite{aas_2018}
$m^*=\pm\hbar^2 E_g/[2(f_2 a)^2]$. Since $f_4<0$ (cf. Table~\ref{tableII}), a tensile hydrostatic strain $(\varepsilon_H>0)$
decreases the $E_g$. Hence, this decrease in band gap is the common origin for the improvement in both BC and OMM. As applying a tensile
strain to a monolayer TMD is far less problematic than a compressive one which would lead to the buckling of the membrane,\cite{amorim_2016} 
it warrants a realistic strain enhancement of the geometric band properties.  

\section{\label{sec:Conclusions}Conclusions\\}
The appealing features of TMDs can be traced down largely to geometric band effects controlled by BC and OMM. Moreover, 
they can be widely tuned by exerting strain. To harness these in device applications accurate and physically-transparent 
band structure tools are needed. In this work we offer two options: a $k\cdot p$ model (two- or a four-band) having 
an up-to-date parameter set, and a strained extension of a three-band TB Hamiltonian. Despite their simplicity, 
both capture the essential physics that govern the variation of BC and OMM, but with different validity ranges 
around the $K$ valley. Quantitatively, we report under reasonable biaxial tensile strains (about 2.5\%) that these 
can be doubled in value. It is straightforward to incorporate excitonic effects to this framework. \cite{aas_2018} 
Thus, these models may serve for TMD device modeling purposes under electric, magnetic or optical excitations 
in addition to strain. 

\providecommand{\noopsort}[1]{}\providecommand{\singleletter}[1]{#1}%


\begin{thebibliography}{46}%
\makeatletter
\providecommand \@ifxundefined [1]{%
 \@ifx{#1\undefined}
}%
\providecommand \@ifnum [1]{%
 \ifnum #1\expandafter \@firstoftwo
 \else \expandafter \@secondoftwo
 \fi
}%
\providecommand \@ifx [1]{%
 \ifx #1\expandafter \@firstoftwo
 \else \expandafter \@secondoftwo
 \fi
}%
\providecommand \natexlab [1]{#1}%
\providecommand \enquote  [1]{``#1''}%
\providecommand \bibnamefont  [1]{#1}%
\providecommand \bibfnamefont [1]{#1}%
\providecommand \citenamefont [1]{#1}%
\providecommand \href@noop [0]{\@secondoftwo}%
\providecommand \href [0]{\begingroup \@sanitize@url \@href}%
\providecommand \@href[1]{\@@startlink{#1}\@@href}%
\providecommand \@@href[1]{\endgroup#1\@@endlink}%
\providecommand \@sanitize@url [0]{\catcode `\\12\catcode `\$12\catcode
  `\&12\catcode `\#12\catcode `\^12\catcode `\_12\catcode `\%12\relax}%
\providecommand \@@startlink[1]{}%
\providecommand \@@endlink[0]{}%
\providecommand \url  [0]{\begingroup\@sanitize@url \@url }%
\providecommand \@url [1]{\endgroup\@href {#1}{\urlprefix }}%
\providecommand \urlprefix  [0]{URL }%
\providecommand \Eprint [0]{\href }%
\providecommand \doibase [0]{http://dx.doi.org/}%
\providecommand \selectlanguage [0]{\@gobble}%
\providecommand \bibinfo  [0]{\@secondoftwo}%
\providecommand \bibfield  [0]{\@secondoftwo}%
\providecommand \translation [1]{[#1]}%
\providecommand \BibitemOpen [0]{}%
\providecommand \bibitemStop [0]{}%
\providecommand \bibitemNoStop [0]{.\EOS\space}%
\providecommand \EOS [0]{\spacefactor3000\relax}%
\providecommand \BibitemShut  [1]{\csname bibitem#1\endcsname}%
\let\auto@bib@innerbib\@empty
\bibitem [{\citenamefont {Xu}\ \emph {et~al.}(2014)\citenamefont {Xu},
  \citenamefont {Yao}, \citenamefont {Xiao},\ and\ \citenamefont
  {Heinz}}]{xu_2014}%
  \BibitemOpen
  \bibfield  {author} {\bibinfo {author} {\bibfnamefont {X.}~\bibnamefont
  {Xu}}, \bibinfo {author} {\bibfnamefont {W.}~\bibnamefont {Yao}}, \bibinfo
  {author} {\bibfnamefont {D.}~\bibnamefont {Xiao}}, \ and\ \bibinfo {author}
  {\bibfnamefont {T.~F.}\ \bibnamefont {Heinz}},\ }\href@noop {} {\bibfield
  {journal} {\bibinfo  {journal} {Nat. Phys.}\ }\textbf {\bibinfo {volume}
  {10}},\ \bibinfo {pages} {343--350} (\bibinfo {year} {2014})}\BibitemShut
  {NoStop}%
\bibitem [{\citenamefont {Manzeli}\ \emph {et~al.}(2017)\citenamefont
  {Manzeli}, \citenamefont {Ovchinnikov}, \citenamefont {Pasquier},
  \citenamefont {Yazyev},\ and\ \citenamefont {Kis}}]{manzeli_2017}%
  \BibitemOpen
  \bibfield  {author} {\bibinfo {author} {\bibfnamefont {S.}~\bibnamefont
  {Manzeli}}, \bibinfo {author} {\bibfnamefont {D.}~\bibnamefont
  {Ovchinnikov}}, \bibinfo {author} {\bibfnamefont {D.}~\bibnamefont
  {Pasquier}}, \bibinfo {author} {\bibfnamefont {O.~V.}\ \bibnamefont
  {Yazyev}}, \ and\ \bibinfo {author} {\bibfnamefont {A.}~\bibnamefont {Kis}},\
  }\href@noop {} {\bibfield  {journal} {\bibinfo  {journal} {Nat. Rev. Mater.}\
  }\textbf {\bibinfo {volume} {2}},\ \bibinfo {pages} {17033} (\bibinfo {year}
  {2017})}\BibitemShut {NoStop}%
\bibitem [{\citenamefont {Mak}, \citenamefont {Xiao},\ and\ \citenamefont
  {Shan}(2018)}]{mak_2018}%
  \BibitemOpen
  \bibfield  {author} {\bibinfo {author} {\bibfnamefont {K.~F.}\ \bibnamefont
  {Mak}}, \bibinfo {author} {\bibfnamefont {D.}~\bibnamefont {Xiao}}, \ and\
  \bibinfo {author} {\bibfnamefont {J.}~\bibnamefont {Shan}},\ }\href@noop {}
  {\bibfield  {journal} {\bibinfo  {journal} {Nat. Photon.}\ }\textbf {\bibinfo
  {volume} {12}},\ \bibinfo {pages} {451--460} (\bibinfo {year}
  {2018})}\BibitemShut {NoStop}%
\bibitem [{\citenamefont {Yamamoto}\ \emph {et~al.}(2015)\citenamefont
  {Yamamoto}, \citenamefont {Shimazaki}, \citenamefont {Borzenets},\ and\
  \citenamefont {Tarucha}}]{yamamoto_2015}%
  \BibitemOpen
  \bibfield  {author} {\bibinfo {author} {\bibfnamefont {M.}~\bibnamefont
  {Yamamoto}}, \bibinfo {author} {\bibfnamefont {Y.}~\bibnamefont {Shimazaki}},
  \bibinfo {author} {\bibfnamefont {I.~V.}\ \bibnamefont {Borzenets}}, \ and\
  \bibinfo {author} {\bibfnamefont {S.}~\bibnamefont {Tarucha}},\ }\href@noop
  {} {\bibfield  {journal} {\bibinfo  {journal} {J. Phys. Soc. Jpn.}\ }\textbf
  {\bibinfo {volume} {84}},\ \bibinfo {pages} {121006} (\bibinfo {year}
  {2015})}\BibitemShut {NoStop}%
\bibitem [{\citenamefont {Schaibley}\ \emph {et~al.}(2016)\citenamefont
  {Schaibley}, \citenamefont {Yu}, \citenamefont {Clark}, \citenamefont
  {Rivera}, \citenamefont {Ross}, \citenamefont {Seyler}, \citenamefont {Yao},\
  and\ \citenamefont {Xu}}]{schaibley_2016}%
  \BibitemOpen
  \bibfield  {author} {\bibinfo {author} {\bibfnamefont {J.~R.}\ \bibnamefont
  {Schaibley}}, \bibinfo {author} {\bibfnamefont {H.}~\bibnamefont {Yu}},
  \bibinfo {author} {\bibfnamefont {G.}~\bibnamefont {Clark}}, \bibinfo
  {author} {\bibfnamefont {P.}~\bibnamefont {Rivera}}, \bibinfo {author}
  {\bibfnamefont {J.~S.}\ \bibnamefont {Ross}}, \bibinfo {author}
  {\bibfnamefont {K.~L.}\ \bibnamefont {Seyler}}, \bibinfo {author}
  {\bibfnamefont {W.}~\bibnamefont {Yao}}, \ and\ \bibinfo {author}
  {\bibfnamefont {X.}~\bibnamefont {Xu}},\ }\href@noop {} {\bibfield  {journal}
  {\bibinfo  {journal} {Nat. Rev. Mat.}\ }\textbf {\bibinfo {volume} {1}},\
  \bibinfo {pages} {16055} (\bibinfo {year} {2016})}\BibitemShut {NoStop}%
\bibitem [{\citenamefont {Xiao}, \citenamefont {Yao},\ and\ \citenamefont
  {Niu}(2007)}]{xiao_2007}%
  \BibitemOpen
  \bibfield  {author} {\bibinfo {author} {\bibfnamefont {D.}~\bibnamefont
  {Xiao}}, \bibinfo {author} {\bibfnamefont {W.}~\bibnamefont {Yao}}, \ and\
  \bibinfo {author} {\bibfnamefont {Q.}~\bibnamefont {Niu}},\ }\href@noop {}
  {\bibfield  {journal} {\bibinfo  {journal} {Phys. Rev. Lett.}\ }\textbf
  {\bibinfo {volume} {99}},\ \bibinfo {pages} {236809} (\bibinfo {year}
  {2007})}\BibitemShut {NoStop}%
\bibitem [{\citenamefont {Yao}, \citenamefont {Xiao},\ and\ \citenamefont
  {Niu}(2008)}]{yao_2008}%
  \BibitemOpen
  \bibfield  {author} {\bibinfo {author} {\bibfnamefont {W.}~\bibnamefont
  {Yao}}, \bibinfo {author} {\bibfnamefont {D.}~\bibnamefont {Xiao}}, \ and\
  \bibinfo {author} {\bibfnamefont {Q.}~\bibnamefont {Niu}},\ }\href@noop {}
  {\bibfield  {journal} {\bibinfo  {journal} {Phys. Rev. B}\ }\textbf {\bibinfo
  {volume} {77}},\ \bibinfo {pages} {235406} (\bibinfo {year}
  {2008})}\BibitemShut {NoStop}%
\bibitem [{\citenamefont {Xiao}\ \emph {et~al.}(2012)\citenamefont {Xiao},
  \citenamefont {Liu}, \citenamefont {Feng}, \citenamefont {Xu},\ and\
  \citenamefont {Yao}}]{xiao_2012}%
  \BibitemOpen
  \bibfield  {author} {\bibinfo {author} {\bibfnamefont {D.}~\bibnamefont
  {Xiao}}, \bibinfo {author} {\bibfnamefont {G.~B.}\ \bibnamefont {Liu}},
  \bibinfo {author} {\bibfnamefont {W.}~\bibnamefont {Feng}}, \bibinfo {author}
  {\bibfnamefont {X.}~\bibnamefont {Xu}}, \ and\ \bibinfo {author}
  {\bibfnamefont {W.}~\bibnamefont {Yao}},\ }\href@noop {} {\bibfield
  {journal} {\bibinfo  {journal} {Phys. Rev. Lett.}\ }\textbf {\bibinfo
  {volume} {108}},\ \bibinfo {pages} {196802} (\bibinfo {year}
  {2012})}\BibitemShut {NoStop}%
\bibitem [{\citenamefont {Mak}\ \emph {et~al.}(2014)\citenamefont {Mak},
  \citenamefont {McGill}, \citenamefont {Park},\ and\ \citenamefont
  {McEuen}}]{mak_2014}%
  \BibitemOpen
  \bibfield  {author} {\bibinfo {author} {\bibfnamefont {K.~F.}\ \bibnamefont
  {Mak}}, \bibinfo {author} {\bibfnamefont {K.~L.}\ \bibnamefont {McGill}},
  \bibinfo {author} {\bibfnamefont {J.}~\bibnamefont {Park}}, \ and\ \bibinfo
  {author} {\bibfnamefont {P.~L.}\ \bibnamefont {McEuen}},\ }\href@noop {}
  {\bibfield  {journal} {\bibinfo  {journal} {Science}\ }\textbf {\bibinfo
  {volume} {344}},\ \bibinfo {pages} {1489--1492} (\bibinfo {year}
  {2014})}\BibitemShut {NoStop}%
\bibitem [{\citenamefont {Wu}\ \emph {et~al.}(2019)\citenamefont {Wu},
  \citenamefont {Zhou}, \citenamefont {Cai}, \citenamefont {Cheung},
  \citenamefont {Liu}, \citenamefont {Huang}, \citenamefont {Lin},
  \citenamefont {Han}, \citenamefont {An}, \citenamefont {Wang}, \citenamefont
  {Xu}, \citenamefont {Long}, \citenamefont {Cheng}, \citenamefont {Law},
  \citenamefont {Zhang},\ and\ \citenamefont {Wang}}]{wu_2019}%
  \BibitemOpen
  \bibfield  {author} {\bibinfo {author} {\bibfnamefont {Z.}~\bibnamefont
  {Wu}}, \bibinfo {author} {\bibfnamefont {B.~T.}\ \bibnamefont {Zhou}},
  \bibinfo {author} {\bibfnamefont {X.}~\bibnamefont {Cai}}, \bibinfo {author}
  {\bibfnamefont {P.}~\bibnamefont {Cheung}}, \bibinfo {author} {\bibfnamefont
  {G.-B.}\ \bibnamefont {Liu}}, \bibinfo {author} {\bibfnamefont
  {M.}~\bibnamefont {Huang}}, \bibinfo {author} {\bibfnamefont
  {J.}~\bibnamefont {Lin}}, \bibinfo {author} {\bibfnamefont {T.}~\bibnamefont
  {Han}}, \bibinfo {author} {\bibfnamefont {L.}~\bibnamefont {An}}, \bibinfo
  {author} {\bibfnamefont {Y.}~\bibnamefont {Wang}}, \bibinfo {author}
  {\bibfnamefont {S.}~\bibnamefont {Xu}}, \bibinfo {author} {\bibfnamefont
  {G.}~\bibnamefont {Long}}, \bibinfo {author} {\bibfnamefont {C.}~\bibnamefont
  {Cheng}}, \bibinfo {author} {\bibfnamefont {K.~T.}\ \bibnamefont {Law}},
  \bibinfo {author} {\bibfnamefont {F.}~\bibnamefont {Zhang}}, \ and\ \bibinfo
  {author} {\bibfnamefont {N.}~\bibnamefont {Wang}},\ }\href@noop {} {\bibfield
   {journal} {\bibinfo  {journal} {Nat. Comm.}\ }\textbf {\bibinfo {volume}
  {10}},\ \bibinfo {pages} {611} (\bibinfo {year} {2019})}\BibitemShut
  {NoStop}%
\bibitem [{\citenamefont {Cao}\ \emph {et~al.}(2012)\citenamefont {Cao},
  \citenamefont {Wang}, \citenamefont {Han}, \citenamefont {Ye}, \citenamefont
  {Zhu}, \citenamefont {Shi}, \citenamefont {Niu}, \citenamefont {Tan},
  \citenamefont {Wang}, \citenamefont {Liu},\ and\ \citenamefont
  {Feng}}]{cao_2012}%
  \BibitemOpen
  \bibfield  {author} {\bibinfo {author} {\bibfnamefont {T.}~\bibnamefont
  {Cao}}, \bibinfo {author} {\bibfnamefont {G.}~\bibnamefont {Wang}}, \bibinfo
  {author} {\bibfnamefont {W.}~\bibnamefont {Han}}, \bibinfo {author}
  {\bibfnamefont {H.}~\bibnamefont {Ye}}, \bibinfo {author} {\bibfnamefont
  {C.}~\bibnamefont {Zhu}}, \bibinfo {author} {\bibfnamefont {J.}~\bibnamefont
  {Shi}}, \bibinfo {author} {\bibfnamefont {Q.}~\bibnamefont {Niu}}, \bibinfo
  {author} {\bibfnamefont {P.}~\bibnamefont {Tan}}, \bibinfo {author}
  {\bibfnamefont {E.}~\bibnamefont {Wang}}, \bibinfo {author} {\bibfnamefont
  {B.}~\bibnamefont {Liu}}, \ and\ \bibinfo {author} {\bibfnamefont
  {J.}~\bibnamefont {Feng}},\ }\href@noop {} {\bibfield  {journal} {\bibinfo
  {journal} {Nat. Commun.}\ }\textbf {\bibinfo {volume} {3}},\ \bibinfo {pages}
  {887} (\bibinfo {year} {2012})}\BibitemShut {NoStop}%
\bibitem [{\citenamefont {Vanderbilt}(2018)}]{vanderbilt_2018}%
  \BibitemOpen
  \bibfield  {author} {\bibinfo {author} {\bibfnamefont {D.~V.}\ \bibnamefont
  {Vanderbilt}},\ }\href@noop {} {\emph {\bibinfo {title} {Berry Phases in
  Electronic Structure Theory}}}\ (\bibinfo  {publisher} {Cambridge University
  Press, Cambridge},\ \bibinfo {year} {2018})\BibitemShut {NoStop}%
\bibitem [{\citenamefont {Xiao}, \citenamefont {Chang},\ and\ \citenamefont
  {Niu}(2007)}]{xiao_2010}%
  \BibitemOpen
  \bibfield  {author} {\bibinfo {author} {\bibfnamefont {D.}~\bibnamefont
  {Xiao}}, \bibinfo {author} {\bibfnamefont {M.-C.}\ \bibnamefont {Chang}}, \
  and\ \bibinfo {author} {\bibfnamefont {Q.}~\bibnamefont {Niu}},\ }\href@noop
  {} {\bibfield  {journal} {\bibinfo  {journal} {Rev. Mod. Phys.}\ }\textbf
  {\bibinfo {volume} {82}},\ \bibinfo {pages} {1959--2007} (\bibinfo {year}
  {2007})}\BibitemShut {NoStop}%
\bibitem [{\citenamefont {Zhang}, \citenamefont {Shan},\ and\ \citenamefont
  {Xiao}(2018)}]{zhang_2018}%
  \BibitemOpen
  \bibfield  {author} {\bibinfo {author} {\bibfnamefont {X.}~\bibnamefont
  {Zhang}}, \bibinfo {author} {\bibfnamefont {W.-Y.}\ \bibnamefont {Shan}}, \
  and\ \bibinfo {author} {\bibfnamefont {D.}~\bibnamefont {Xiao}},\ }\href@noop
  {} {\bibfield  {journal} {\bibinfo  {journal} {Phys. Rev. Lett.}\ }\textbf
  {\bibinfo {volume} {120}},\ \bibinfo {pages} {077401} (\bibinfo {year}
  {2018})}\BibitemShut {NoStop}%
\bibitem [{\citenamefont {Cao}, \citenamefont {Wu},\ and\ \citenamefont
  {Louie}(2018)}]{cao_2018}%
  \BibitemOpen
  \bibfield  {author} {\bibinfo {author} {\bibfnamefont {T.}~\bibnamefont
  {Cao}}, \bibinfo {author} {\bibfnamefont {M.}~\bibnamefont {Wu}}, \ and\
  \bibinfo {author} {\bibfnamefont {S.~G.}\ \bibnamefont {Louie}},\ }\href@noop
  {} {\bibfield  {journal} {\bibinfo  {journal} {Phys. Rev. Lett.}\ }\textbf
  {\bibinfo {volume} {120}},\ \bibinfo {pages} {077401} (\bibinfo {year}
  {2018})}\BibitemShut {NoStop}%
\bibitem [{\citenamefont {Srivastava}\ and\ \citenamefont
  {Imamo\u{g}lu}(2015)}]{srivastava_2015a}%
  \BibitemOpen
  \bibfield  {author} {\bibinfo {author} {\bibfnamefont {A.}~\bibnamefont
  {Srivastava}}\ and\ \bibinfo {author} {\bibfnamefont {A.}~\bibnamefont
  {Imamo\u{g}lu}},\ }\href@noop {} {\bibfield  {journal} {\bibinfo  {journal}
  {Phys. Rev. Lett.}\ }\textbf {\bibinfo {volume} {115}},\ \bibinfo {pages}
  {166802} (\bibinfo {year} {2015})}\BibitemShut {NoStop}%
\bibitem [{\citenamefont {Zhou}\ \emph {et~al.}(2015)\citenamefont {Zhou},
  \citenamefont {Shan}, \citenamefont {Yao},\ and\ \citenamefont
  {Xiao}}]{zhou_2015}%
  \BibitemOpen
  \bibfield  {author} {\bibinfo {author} {\bibfnamefont {J.}~\bibnamefont
  {Zhou}}, \bibinfo {author} {\bibfnamefont {W.-Y.}\ \bibnamefont {Shan}},
  \bibinfo {author} {\bibfnamefont {W.}~\bibnamefont {Yao}}, \ and\ \bibinfo
  {author} {\bibfnamefont {D.}~\bibnamefont {Xiao}},\ }\href@noop {} {\bibfield
   {journal} {\bibinfo  {journal} {Phys. Rev. Lett.}\ }\textbf {\bibinfo
  {volume} {115}},\ \bibinfo {pages} {166803} (\bibinfo {year}
  {2015})}\BibitemShut {NoStop}%
\bibitem [{\citenamefont {Chang}\ and\ \citenamefont {Niu}(2008)}]{chang_2008}%
  \BibitemOpen
  \bibfield  {author} {\bibinfo {author} {\bibfnamefont {M.-C.}\ \bibnamefont
  {Chang}}\ and\ \bibinfo {author} {\bibfnamefont {Q.}~\bibnamefont {Niu}},\
  }\href@noop {} {\bibfield  {journal} {\bibinfo  {journal} {J. Phys.: Condens.
  Matter}\ }\textbf {\bibinfo {volume} {20}},\ \bibinfo {pages} {193202}
  (\bibinfo {year} {2008})}\BibitemShut {NoStop}%
\bibitem [{\citenamefont {Brooks}\ and\ \citenamefont
  {Burkard}(2017)}]{brooks_2017}%
  \BibitemOpen
  \bibfield  {author} {\bibinfo {author} {\bibfnamefont {M.}~\bibnamefont
  {Brooks}}\ and\ \bibinfo {author} {\bibfnamefont {G.}~\bibnamefont
  {Burkard}},\ }\href@noop {} {\bibfield  {journal} {\bibinfo  {journal} {Phys.
  Rev. B}\ }\textbf {\bibinfo {volume} {95}},\ \bibinfo {pages} {245411}
  (\bibinfo {year} {2017})}\BibitemShut {NoStop}%
\bibitem [{\citenamefont {Rybkovskiy}, \citenamefont {C.Gerber},\ and\
  \citenamefont {Durnev}(2017)}]{rybkovskiy_2017}%
  \BibitemOpen
  \bibfield  {author} {\bibinfo {author} {\bibfnamefont {D.~V.}\ \bibnamefont
  {Rybkovskiy}}, \bibinfo {author} {\bibfnamefont {I.}~\bibnamefont
  {C.Gerber}}, \ and\ \bibinfo {author} {\bibfnamefont {M.~V.}\ \bibnamefont
  {Durnev}},\ }\href@noop {} {\bibfield  {journal} {\bibinfo  {journal} {Phys.
  Rev. B}\ }\textbf {\bibinfo {volume} {95}},\ \bibinfo {pages} {155406}
  (\bibinfo {year} {2017})}\BibitemShut {NoStop}%
\bibitem [{\citenamefont {Aivazian}\ \emph {et~al.}(2015)\citenamefont
  {Aivazian}, \citenamefont {Gong}, \citenamefont {Jones}, \citenamefont {Chu},
  \citenamefont {Yan}, \citenamefont {Mandrus}, \citenamefont {Zhang},
  \citenamefont {Cobden}, \citenamefont {Yao},\ and\ \citenamefont
  {Xu}}]{aivazian_2015}%
  \BibitemOpen
  \bibfield  {author} {\bibinfo {author} {\bibfnamefont {G.}~\bibnamefont
  {Aivazian}}, \bibinfo {author} {\bibfnamefont {Z.}~\bibnamefont {Gong}},
  \bibinfo {author} {\bibfnamefont {A.~M.}\ \bibnamefont {Jones}}, \bibinfo
  {author} {\bibfnamefont {R.~L.}\ \bibnamefont {Chu}}, \bibinfo {author}
  {\bibfnamefont {J.}~\bibnamefont {Yan}}, \bibinfo {author} {\bibfnamefont
  {D.~G.}\ \bibnamefont {Mandrus}}, \bibinfo {author} {\bibfnamefont
  {C.}~\bibnamefont {Zhang}}, \bibinfo {author} {\bibfnamefont
  {D.}~\bibnamefont {Cobden}}, \bibinfo {author} {\bibfnamefont
  {W.}~\bibnamefont {Yao}}, \ and\ \bibinfo {author} {\bibfnamefont
  {X.}~\bibnamefont {Xu}},\ }\href@noop {} {\bibfield  {journal} {\bibinfo
  {journal} {Nat. Phys.}\ }\textbf {\bibinfo {volume} {11}},\ \bibinfo {pages}
  {148--152} (\bibinfo {year} {2015})}\BibitemShut {NoStop}%
\bibitem [{\citenamefont {Srivastava}\ \emph {et~al.}(2015)\citenamefont
  {Srivastava}, \citenamefont {Sidler}, \citenamefont {Allain}, \citenamefont
  {Lembke}, \citenamefont {Kis},\ and\ \citenamefont
  {Imamo\u{g}lu}}]{srivastava_2015b}%
  \BibitemOpen
  \bibfield  {author} {\bibinfo {author} {\bibfnamefont {A.}~\bibnamefont
  {Srivastava}}, \bibinfo {author} {\bibfnamefont {M.}~\bibnamefont {Sidler}},
  \bibinfo {author} {\bibfnamefont {A.~V.}\ \bibnamefont {Allain}}, \bibinfo
  {author} {\bibfnamefont {D.~S.}\ \bibnamefont {Lembke}}, \bibinfo {author}
  {\bibfnamefont {A.}~\bibnamefont {Kis}}, \ and\ \bibinfo {author}
  {\bibfnamefont {A.}~\bibnamefont {Imamo\u{g}lu}},\ }\href@noop {} {\bibfield
  {journal} {\bibinfo  {journal} {Nat. Phys.}\ }\textbf {\bibinfo {volume}
  {11}},\ \bibinfo {pages} {141--147} (\bibinfo {year} {2015})}\BibitemShut
  {NoStop}%
\bibitem [{\citenamefont {Sch{\"{u}}ler}\ \emph {et~al.}(2019)\citenamefont
  {Sch{\"{u}}ler}, \citenamefont {Giovannini}, \citenamefont {H{\"{u}}bener},
  \citenamefont {Rubio}, \citenamefont {Sentef},\ and\ \citenamefont
  {Werner}}]{schuller_2019}%
  \BibitemOpen
  \bibfield  {author} {\bibinfo {author} {\bibfnamefont {M.}~\bibnamefont
  {Sch{\"{u}}ler}}, \bibinfo {author} {\bibfnamefont {U.~D.}\ \bibnamefont
  {Giovannini}}, \bibinfo {author} {\bibfnamefont {H.}~\bibnamefont
  {H{\"{u}}bener}}, \bibinfo {author} {\bibfnamefont {A.}~\bibnamefont
  {Rubio}}, \bibinfo {author} {\bibfnamefont {M.~A.}\ \bibnamefont {Sentef}}, \
  and\ \bibinfo {author} {\bibfnamefont {P.}~\bibnamefont {Werner}},\
  }\href@noop {} {\bibfield  {journal} {\bibinfo  {journal} {arXiv preprint
  arXiv:1905.09404}\ } (\bibinfo {year} {2019})}\BibitemShut {NoStop}%
\bibitem [{\citenamefont {Bertolazzi}, \citenamefont {Brivio},\ and\
  \citenamefont {Kis}(2011)}]{bertolazzi_2011}%
  \BibitemOpen
  \bibfield  {author} {\bibinfo {author} {\bibfnamefont {S.}~\bibnamefont
  {Bertolazzi}}, \bibinfo {author} {\bibfnamefont {J.}~\bibnamefont {Brivio}},
  \ and\ \bibinfo {author} {\bibfnamefont {A.}~\bibnamefont {Kis}},\
  }\href@noop {} {\bibfield  {journal} {\bibinfo  {journal} {ACS Nano}\
  }\textbf {\bibinfo {volume} {5}},\ \bibinfo {pages} {9703--9709} (\bibinfo
  {year} {2011})}\BibitemShut {NoStop}%
\bibitem [{\citenamefont {Korm{\'{a}}nyos}\ \emph {et~al.}(2015)\citenamefont
  {Korm{\'{a}}nyos}, \citenamefont {Burkard}, \citenamefont {Gmitra},
  \citenamefont {Fabian}, \citenamefont {Z{\'{o}}lyomi}, \citenamefont
  {Drummond},\ and\ \citenamefont {Fal'ko}}]{kormanyos_2015}%
  \BibitemOpen
  \bibfield  {author} {\bibinfo {author} {\bibfnamefont {A.}~\bibnamefont
  {Korm{\'{a}}nyos}}, \bibinfo {author} {\bibfnamefont {G.}~\bibnamefont
  {Burkard}}, \bibinfo {author} {\bibfnamefont {M.}~\bibnamefont {Gmitra}},
  \bibinfo {author} {\bibfnamefont {J.}~\bibnamefont {Fabian}}, \bibinfo
  {author} {\bibfnamefont {V.}~\bibnamefont {Z{\'{o}}lyomi}}, \bibinfo {author}
  {\bibfnamefont {N.~D.}\ \bibnamefont {Drummond}}, \ and\ \bibinfo {author}
  {\bibfnamefont {V.~I.}\ \bibnamefont {Fal'ko}},\ }\href@noop {} {\bibfield
  {journal} {\bibinfo  {journal} {2D Materials}\ }\textbf {\bibinfo {volume}
  {2}},\ \bibinfo {pages} {022001} (\bibinfo {year} {2015})}\BibitemShut
  {NoStop}%
\bibitem [{\citenamefont {Rostami}\ \emph {et~al.}(2015)\citenamefont
  {Rostami}, \citenamefont {Roldán}, \citenamefont {Cappelluti}, \citenamefont
  {Asgari},\ and\ \citenamefont {Guinea}}]{rostami_2015}%
  \BibitemOpen
  \bibfield  {author} {\bibinfo {author} {\bibfnamefont {H.}~\bibnamefont
  {Rostami}}, \bibinfo {author} {\bibfnamefont {R.}~\bibnamefont {Roldán}},
  \bibinfo {author} {\bibfnamefont {E.}~\bibnamefont {Cappelluti}}, \bibinfo
  {author} {\bibfnamefont {R.}~\bibnamefont {Asgari}}, \ and\ \bibinfo {author}
  {\bibfnamefont {F.}~\bibnamefont {Guinea}},\ }\href@noop {} {\bibfield
  {journal} {\bibinfo  {journal} {Phys. Rev. B}\ }\textbf {\bibinfo {volume}
  {92}},\ \bibinfo {pages} {195402} (\bibinfo {year} {2015})}\BibitemShut
  {NoStop}%
\bibitem [{\citenamefont {Pearce}, \citenamefont {Mariani},\ and\ \citenamefont
  {Burkard}(2016)}]{pearce_2016}%
  \BibitemOpen
  \bibfield  {author} {\bibinfo {author} {\bibfnamefont {A.~J.}\ \bibnamefont
  {Pearce}}, \bibinfo {author} {\bibfnamefont {E.}~\bibnamefont {Mariani}}, \
  and\ \bibinfo {author} {\bibfnamefont {G.}~\bibnamefont {Burkard}},\
  }\href@noop {} {\bibfield  {journal} {\bibinfo  {journal} {Phys. Rev. B}\
  }\textbf {\bibinfo {volume} {94}},\ \bibinfo {pages} {155416} (\bibinfo
  {year} {2016})}\BibitemShut {NoStop}%
\bibitem [{\citenamefont {Sevik}\ \emph {et~al.}(2017)\citenamefont {Sevik},
  \citenamefont {Wallbank}, \citenamefont {Gülseren}, \citenamefont
  {Peeters},\ and\ \citenamefont {Çakır}}]{sevik_2017}%
  \BibitemOpen
  \bibfield  {author} {\bibinfo {author} {\bibfnamefont {C.}~\bibnamefont
  {Sevik}}, \bibinfo {author} {\bibfnamefont {J.~R.}\ \bibnamefont {Wallbank}},
  \bibinfo {author} {\bibfnamefont {O.}~\bibnamefont {Gülseren}}, \bibinfo
  {author} {\bibfnamefont {F.~M.}\ \bibnamefont {Peeters}}, \ and\ \bibinfo
  {author} {\bibfnamefont {D.}~\bibnamefont {Çakır}},\ }\href@noop {}
  {\bibfield  {journal} {\bibinfo  {journal} {2D Mater.}\ }\textbf {\bibinfo
  {volume} {4}},\ \bibinfo {pages} {035025} (\bibinfo {year}
  {2017})}\BibitemShut {NoStop}%
\bibitem [{\citenamefont {Fang}\ \emph
  {et~al.}(2018{\natexlab{a}})\citenamefont {Fang}, \citenamefont {Carr},
  \citenamefont {Cazalilla},\ and\ \citenamefont {Kaxiras}}]{fang_2018}%
  \BibitemOpen
  \bibfield  {author} {\bibinfo {author} {\bibfnamefont {S.}~\bibnamefont
  {Fang}}, \bibinfo {author} {\bibfnamefont {S.}~\bibnamefont {Carr}}, \bibinfo
  {author} {\bibfnamefont {M.~A.}\ \bibnamefont {Cazalilla}}, \ and\ \bibinfo
  {author} {\bibfnamefont {E.}~\bibnamefont {Kaxiras}},\ }\href@noop {}
  {\bibfield  {journal} {\bibinfo  {journal} {Phys. Rev. B}\ }\textbf {\bibinfo
  {volume} {98}},\ \bibinfo {pages} {075106} (\bibinfo {year}
  {2018}{\natexlab{a}})}\BibitemShut {NoStop}%
\bibitem [{\citenamefont {Aas}\ and\ \citenamefont {Bulutay}(2018)}]{aas_2018}%
  \BibitemOpen
  \bibfield  {author} {\bibinfo {author} {\bibfnamefont {S.}~\bibnamefont
  {Aas}}\ and\ \bibinfo {author} {\bibfnamefont {C.}~\bibnamefont {Bulutay}},\
  }\href@noop {} {\bibfield  {journal} {\bibinfo  {journal} {Opt. Express}\
  }\textbf {\bibinfo {volume} {26}},\ \bibinfo {pages} {28672--28681} (\bibinfo
  {year} {2018})}\BibitemShut {NoStop}%
\bibitem [{\citenamefont {Zhang}\ \emph {et~al.}(2014)\citenamefont {Zhang},
  \citenamefont {Oka}, \citenamefont {Suzuki}, \citenamefont {Ye},\ and\
  \citenamefont {Iwasa}}]{zhang_2014}%
  \BibitemOpen
  \bibfield  {author} {\bibinfo {author} {\bibfnamefont {Y.~J.}\ \bibnamefont
  {Zhang}}, \bibinfo {author} {\bibfnamefont {T.}~\bibnamefont {Oka}}, \bibinfo
  {author} {\bibfnamefont {R.}~\bibnamefont {Suzuki}}, \bibinfo {author}
  {\bibfnamefont {J.~T.}\ \bibnamefont {Ye}}, \ and\ \bibinfo {author}
  {\bibfnamefont {Y.}~\bibnamefont {Iwasa}},\ }\href@noop {} {\bibfield
  {journal} {\bibinfo  {journal} {Science}\ }\textbf {\bibinfo {volume}
  {344}},\ \bibinfo {pages} {725--728} (\bibinfo {year} {2014})}\BibitemShut
  {NoStop}%
\bibitem [{\citenamefont {Li}\ \emph {et~al.}(2014)\citenamefont {Li},
  \citenamefont {Ludwig}, \citenamefont {Low}, \citenamefont {Chernikov},
  \citenamefont {Cui}, \citenamefont {Arefe}, \citenamefont {Kim},
  \citenamefont {van~der Zande}, \citenamefont {Rigosi}, \citenamefont {Hill},
  \citenamefont {Kim}, \citenamefont {Hone}, \citenamefont {Li}, \citenamefont
  {Smirnov},\ and\ \citenamefont {Heinz}}]{li_2014}%
  \BibitemOpen
  \bibfield  {author} {\bibinfo {author} {\bibfnamefont {Y.}~\bibnamefont
  {Li}}, \bibinfo {author} {\bibfnamefont {J.}~\bibnamefont {Ludwig}}, \bibinfo
  {author} {\bibfnamefont {T.}~\bibnamefont {Low}}, \bibinfo {author}
  {\bibfnamefont {A.}~\bibnamefont {Chernikov}}, \bibinfo {author}
  {\bibfnamefont {X.}~\bibnamefont {Cui}}, \bibinfo {author} {\bibfnamefont
  {G.}~\bibnamefont {Arefe}}, \bibinfo {author} {\bibfnamefont {Y.~D.}\
  \bibnamefont {Kim}}, \bibinfo {author} {\bibfnamefont {A.~M.}\ \bibnamefont
  {van~der Zande}}, \bibinfo {author} {\bibfnamefont {A.}~\bibnamefont
  {Rigosi}}, \bibinfo {author} {\bibfnamefont {H.~M.}\ \bibnamefont {Hill}},
  \bibinfo {author} {\bibfnamefont {S.~H.}\ \bibnamefont {Kim}}, \bibinfo
  {author} {\bibfnamefont {J.}~\bibnamefont {Hone}}, \bibinfo {author}
  {\bibfnamefont {Z.}~\bibnamefont {Li}}, \bibinfo {author} {\bibfnamefont
  {D.}~\bibnamefont {Smirnov}}, \ and\ \bibinfo {author} {\bibfnamefont
  {T.~F.}\ \bibnamefont {Heinz}},\ }\href@noop {} {\bibfield  {journal}
  {\bibinfo  {journal} {Phys. Rev. Lett.}\ }\textbf {\bibinfo {volume} {113}},\
  \bibinfo {pages} {266804} (\bibinfo {year} {2014})}\BibitemShut {NoStop}%
\bibitem [{\citenamefont {MacNeill}\ \emph {et~al.}(2015)\citenamefont
  {MacNeill}, \citenamefont {Heikes}, \citenamefont {Mak}, \citenamefont
  {Anderson}, \citenamefont {Korm{\'{a}}nyos}, \citenamefont {Z{\'{o}}lyomi},
  \citenamefont {Park},\ and\ \citenamefont {Ralph}}]{macneill_2015}%
  \BibitemOpen
  \bibfield  {author} {\bibinfo {author} {\bibfnamefont {D.}~\bibnamefont
  {MacNeill}}, \bibinfo {author} {\bibfnamefont {C.}~\bibnamefont {Heikes}},
  \bibinfo {author} {\bibfnamefont {K.~F.}\ \bibnamefont {Mak}}, \bibinfo
  {author} {\bibfnamefont {Z.}~\bibnamefont {Anderson}}, \bibinfo {author}
  {\bibfnamefont {A.}~\bibnamefont {Korm{\'{a}}nyos}}, \bibinfo {author}
  {\bibfnamefont {V.}~\bibnamefont {Z{\'{o}}lyomi}}, \bibinfo {author}
  {\bibfnamefont {J.}~\bibnamefont {Park}}, \ and\ \bibinfo {author}
  {\bibfnamefont {D.~C.}\ \bibnamefont {Ralph}},\ }\href@noop {} {\bibfield
  {journal} {\bibinfo  {journal} {Phys. Rev. Lett.}\ }\textbf {\bibinfo
  {volume} {114}},\ \bibinfo {pages} {037401} (\bibinfo {year}
  {2015})}\BibitemShut {NoStop}%
\bibitem [{\citenamefont {Korm{\'{a}}nyos}\ \emph {et~al.}(2013)\citenamefont
  {Korm{\'{a}}nyos}, \citenamefont {Z{\'{o}}lyomi}, \citenamefont {Drummond},
  \citenamefont {Rakyta}, \citenamefont {Burkard},\ and\ \citenamefont
  {Fal'ko}}]{kormanyos_2013}%
  \BibitemOpen
  \bibfield  {author} {\bibinfo {author} {\bibfnamefont {A.}~\bibnamefont
  {Korm{\'{a}}nyos}}, \bibinfo {author} {\bibfnamefont {V.}~\bibnamefont
  {Z{\'{o}}lyomi}}, \bibinfo {author} {\bibfnamefont {N.~D.}\ \bibnamefont
  {Drummond}}, \bibinfo {author} {\bibfnamefont {P.}~\bibnamefont {Rakyta}},
  \bibinfo {author} {\bibfnamefont {G.}~\bibnamefont {Burkard}}, \ and\
  \bibinfo {author} {\bibfnamefont {V.~I.}\ \bibnamefont {Fal'ko}},\
  }\href@noop {} {\bibfield  {journal} {\bibinfo  {journal} {Phys. Rev. B}\
  }\textbf {\bibinfo {volume} {88}},\ \bibinfo {pages} {045416} (\bibinfo
  {year} {2013})}\BibitemShut {NoStop}%
\bibitem [{\citenamefont {Chen}\ \emph {et~al.}(2019)\citenamefont {Chen},
  \citenamefont {Lu}, \citenamefont {Goldstein}, \citenamefont {Tong},
  \citenamefont {Chaves}, \citenamefont {Kunstmann}, \citenamefont
  {Cavalcante}, \citenamefont {Woźniak}, \citenamefont {Seifert},
  \citenamefont {Reichman}, \citenamefont {Taniguchi}, \citenamefont
  {Watanabe}, \citenamefont {Smirnov},\ and\ \citenamefont {Yan}}]{chen_2019}%
  \BibitemOpen
  \bibfield  {author} {\bibinfo {author} {\bibfnamefont {S.-Y.}\ \bibnamefont
  {Chen}}, \bibinfo {author} {\bibfnamefont {Z.}~\bibnamefont {Lu}}, \bibinfo
  {author} {\bibfnamefont {T.}~\bibnamefont {Goldstein}}, \bibinfo {author}
  {\bibfnamefont {J.}~\bibnamefont {Tong}}, \bibinfo {author} {\bibfnamefont
  {A.}~\bibnamefont {Chaves}}, \bibinfo {author} {\bibfnamefont
  {J.}~\bibnamefont {Kunstmann}}, \bibinfo {author} {\bibfnamefont {L.~S.~R.}\
  \bibnamefont {Cavalcante}}, \bibinfo {author} {\bibfnamefont
  {T.}~\bibnamefont {Woźniak}}, \bibinfo {author} {\bibfnamefont
  {G.}~\bibnamefont {Seifert}}, \bibinfo {author} {\bibfnamefont {D.~R.}\
  \bibnamefont {Reichman}}, \bibinfo {author} {\bibfnamefont {T.}~\bibnamefont
  {Taniguchi}}, \bibinfo {author} {\bibfnamefont {K.}~\bibnamefont {Watanabe}},
  \bibinfo {author} {\bibfnamefont {D.}~\bibnamefont {Smirnov}}, \ and\
  \bibinfo {author} {\bibfnamefont {J.}~\bibnamefont {Yan}},\ }\href@noop {}
  {\bibfield  {journal} {\bibinfo  {journal} {Nano Lett.}\ }\textbf {\bibinfo
  {volume} {19}},\ \bibinfo {pages} {2464--2471} (\bibinfo {year}
  {2019})}\BibitemShut {NoStop}%
\bibitem [{\citenamefont {Bromley}, \citenamefont {Murray},\ and\ \citenamefont
  {Yoffe}(1972)}]{bromley_1972}%
  \BibitemOpen
  \bibfield  {author} {\bibinfo {author} {\bibfnamefont {R.~A.}\ \bibnamefont
  {Bromley}}, \bibinfo {author} {\bibfnamefont {R.~B.}\ \bibnamefont {Murray}},
  \ and\ \bibinfo {author} {\bibfnamefont {A.~D.}\ \bibnamefont {Yoffe}},\
  }\href@noop {} {\bibfield  {journal} {\bibinfo  {journal} {J. Phys. C: Solid
  State Phys.}\ }\textbf {\bibinfo {volume} {5}},\ \bibinfo {pages} {759--778}
  (\bibinfo {year} {1972})}\BibitemShut {NoStop}%
\bibitem [{\citenamefont {Cappelluti}\ \emph {et~al.}(2013)\citenamefont
  {Cappelluti}, \citenamefont {Roldán}, \citenamefont {Silva-Guillén},
  \citenamefont {Ordejón},\ and\ \citenamefont {Guinea}}]{cappelluti_2013}%
  \BibitemOpen
  \bibfield  {author} {\bibinfo {author} {\bibfnamefont {E.}~\bibnamefont
  {Cappelluti}}, \bibinfo {author} {\bibfnamefont {R.}~\bibnamefont {Roldán}},
  \bibinfo {author} {\bibfnamefont {A.}~\bibnamefont {Silva-Guillén}},
  \bibinfo {author} {\bibfnamefont {P.}~\bibnamefont {Ordejón}}, \ and\
  \bibinfo {author} {\bibfnamefont {F.}~\bibnamefont {Guinea}},\ }\href@noop {}
  {\bibfield  {journal} {\bibinfo  {journal} {Phys. Rev. B}\ }\textbf {\bibinfo
  {volume} {88}},\ \bibinfo {pages} {075409} (\bibinfo {year}
  {2013})}\BibitemShut {NoStop}%
\bibitem [{\citenamefont {Liu}\ \emph {et~al.}(2013)\citenamefont {Liu},
  \citenamefont {Shan}, \citenamefont {Yao}, \citenamefont {Yao},\ and\
  \citenamefont {Xiao}}]{liu_2013}%
  \BibitemOpen
  \bibfield  {author} {\bibinfo {author} {\bibfnamefont {G.~B.}\ \bibnamefont
  {Liu}}, \bibinfo {author} {\bibfnamefont {W.~Y.}\ \bibnamefont {Shan}},
  \bibinfo {author} {\bibfnamefont {Y.}~\bibnamefont {Yao}}, \bibinfo {author}
  {\bibfnamefont {W.}~\bibnamefont {Yao}}, \ and\ \bibinfo {author}
  {\bibfnamefont {D.}~\bibnamefont {Xiao}},\ }\href@noop {} {\bibfield
  {journal} {\bibinfo  {journal} {Phys. Rev. B}\ }\textbf {\bibinfo {volume}
  {88}},\ \bibinfo {pages} {085433} (\bibinfo {year} {2013})}\BibitemShut
  {NoStop}%
\bibitem [{\citenamefont {Fang}\ \emph
  {et~al.}(2018{\natexlab{b}})\citenamefont {Fang}, \citenamefont {Defo},
  \citenamefont {Shirodkar}, \citenamefont {Lieu}, \citenamefont {Tritsaris},\
  and\ \citenamefont {Kaxiras}}]{fang_2015}%
  \BibitemOpen
  \bibfield  {author} {\bibinfo {author} {\bibfnamefont {S.}~\bibnamefont
  {Fang}}, \bibinfo {author} {\bibfnamefont {R.~K.}\ \bibnamefont {Defo}},
  \bibinfo {author} {\bibfnamefont {S.~N.}\ \bibnamefont {Shirodkar}}, \bibinfo
  {author} {\bibfnamefont {S.}~\bibnamefont {Lieu}}, \bibinfo {author}
  {\bibfnamefont {G.~A.}\ \bibnamefont {Tritsaris}}, \ and\ \bibinfo {author}
  {\bibfnamefont {E.}~\bibnamefont {Kaxiras}},\ }\href@noop {} {\bibfield
  {journal} {\bibinfo  {journal} {Phys. Rev. B}\ }\textbf {\bibinfo {volume}
  {92}},\ \bibinfo {pages} {205108} (\bibinfo {year}
  {2018}{\natexlab{b}})}\BibitemShut {NoStop}%
\bibitem [{\citenamefont {Jeong}\ \emph {et~al.}(2018)\citenamefont {Jeong},
  \citenamefont {Choi}, \citenamefont {Jeong}, \citenamefont {Park},
  \citenamefont {Kim},\ and\ \citenamefont {Cho}}]{jeong_2018}%
  \BibitemOpen
  \bibfield  {author} {\bibinfo {author} {\bibfnamefont {J.}~\bibnamefont
  {Jeong}}, \bibinfo {author} {\bibfnamefont {Y.~H.}\ \bibnamefont {Choi}},
  \bibinfo {author} {\bibfnamefont {K.}~\bibnamefont {Jeong}}, \bibinfo
  {author} {\bibfnamefont {H.}~\bibnamefont {Park}}, \bibinfo {author}
  {\bibfnamefont {D.}~\bibnamefont {Kim}}, \ and\ \bibinfo {author}
  {\bibfnamefont {M.~H.}\ \bibnamefont {Cho}},\ }\href@noop {} {\bibfield
  {journal} {\bibinfo  {journal} {Phys. Rev. B}\ }\textbf {\bibinfo {volume}
  {97}},\ \bibinfo {pages} {075433} (\bibinfo {year} {2018})}\BibitemShut
  {NoStop}%
\bibitem [{\citenamefont {Zhang}\ \emph {et~al.}(2017)\citenamefont {Zhang},
  \citenamefont {Huang}, \citenamefont {Zhao}, \citenamefont {Zhu},
  \citenamefont {Yao}, \citenamefont {Zhou}, \citenamefont {Du},\ and\
  \citenamefont {Xu}}]{zhang_2017}%
  \BibitemOpen
  \bibfield  {author} {\bibinfo {author} {\bibfnamefont {L.}~\bibnamefont
  {Zhang}}, \bibinfo {author} {\bibfnamefont {Y.}~\bibnamefont {Huang}},
  \bibinfo {author} {\bibfnamefont {Q.}~\bibnamefont {Zhao}}, \bibinfo {author}
  {\bibfnamefont {L.}~\bibnamefont {Zhu}}, \bibinfo {author} {\bibfnamefont
  {Z.}~\bibnamefont {Yao}}, \bibinfo {author} {\bibfnamefont {Y.}~\bibnamefont
  {Zhou}}, \bibinfo {author} {\bibfnamefont {W.}~\bibnamefont {Du}}, \ and\
  \bibinfo {author} {\bibfnamefont {X.}~\bibnamefont {Xu}},\ }\href@noop {}
  {\bibfield  {journal} {\bibinfo  {journal} {Phys. Rev. B}\ }\textbf {\bibinfo
  {volume} {96}},\ \bibinfo {pages} {155202} (\bibinfo {year}
  {2017})}\BibitemShut {NoStop}%
\bibitem [{\citenamefont {Rasmussen}\ and\ \citenamefont
  {Thygesen}(2015)}]{rasmussen_2015}%
  \BibitemOpen
  \bibfield  {author} {\bibinfo {author} {\bibfnamefont {F.~A.}\ \bibnamefont
  {Rasmussen}}\ and\ \bibinfo {author} {\bibfnamefont {K.~S.}\ \bibnamefont
  {Thygesen}},\ }\href@noop {} {\bibfield  {journal} {\bibinfo  {journal} {J.
  Phys. Chem. C}\ }\textbf {\bibinfo {volume} {119}},\ \bibinfo {pages}
  {13169--83} (\bibinfo {year} {2015})}\BibitemShut {NoStop}%
\bibitem [{\citenamefont {Thygesen}(2017)}]{thygesen_2017}%
  \BibitemOpen
  \bibfield  {author} {\bibinfo {author} {\bibfnamefont {K.~S.}\ \bibnamefont
  {Thygesen}},\ }\href@noop {} {\bibfield  {journal} {\bibinfo  {journal} {2D
  Mater.}\ }\textbf {\bibinfo {volume} {4}},\ \bibinfo {pages} {022004}
  (\bibinfo {year} {2017})}\BibitemShut {NoStop}%
\bibitem [{\citenamefont {Feng}\ \emph {et~al.}(2012)\citenamefont {Feng},
  \citenamefont {Yao}, \citenamefont {Zhu}, \citenamefont {Zhou}, \citenamefont
  {Yao},\ and\ \citenamefont {Xiao}}]{feng_2012}%
  \BibitemOpen
  \bibfield  {author} {\bibinfo {author} {\bibfnamefont {W.}~\bibnamefont
  {Feng}}, \bibinfo {author} {\bibfnamefont {Y.}~\bibnamefont {Yao}}, \bibinfo
  {author} {\bibfnamefont {W.}~\bibnamefont {Zhu}}, \bibinfo {author}
  {\bibfnamefont {J.}~\bibnamefont {Zhou}}, \bibinfo {author} {\bibfnamefont
  {W.}~\bibnamefont {Yao}}, \ and\ \bibinfo {author} {\bibfnamefont
  {D.}~\bibnamefont {Xiao}},\ }\href@noop {} {\bibfield  {journal} {\bibinfo
  {journal} {Phys. Rev. B}\ }\textbf {\bibinfo {volume} {86}},\ \bibinfo
  {pages} {165108} (\bibinfo {year} {2012})}\BibitemShut {NoStop}%
\bibitem [{\citenamefont {Wang}\ \emph {et~al.}(2018)\citenamefont {Wang},
  \citenamefont {Chernikov}, \citenamefont {Glazov}, \citenamefont {Heinz},
  \citenamefont {Marie}, \citenamefont {Amand},\ and\ \citenamefont
  {Urbaszek}}]{wang_2018}%
  \BibitemOpen
  \bibfield  {author} {\bibinfo {author} {\bibfnamefont {G.}~\bibnamefont
  {Wang}}, \bibinfo {author} {\bibfnamefont {A.}~\bibnamefont {Chernikov}},
  \bibinfo {author} {\bibfnamefont {M.~M.}\ \bibnamefont {Glazov}}, \bibinfo
  {author} {\bibfnamefont {T.~F.}\ \bibnamefont {Heinz}}, \bibinfo {author}
  {\bibfnamefont {X.}~\bibnamefont {Marie}}, \bibinfo {author} {\bibfnamefont
  {T.}~\bibnamefont {Amand}}, \ and\ \bibinfo {author} {\bibfnamefont
  {B.}~\bibnamefont {Urbaszek}},\ }\href@noop {} {\bibfield  {journal}
  {\bibinfo  {journal} {Rev. Mod. Phys.}\ }\textbf {\bibinfo {volume} {90}},\
  \bibinfo {pages} {021001} (\bibinfo {year} {2018})}\BibitemShut {NoStop}%
\bibitem [{\citenamefont {Amorim}\ \emph {et~al.}(2016)\citenamefont {Amorim},
  \citenamefont {Cortijo}, \citenamefont {de~Juan}, \citenamefont {Grushin},
  \citenamefont {Guinea}, \citenamefont {Gutiérrez-Rubio}, \citenamefont
  {Ochoa}, \citenamefont {Parente}, \citenamefont {Roldán}, \citenamefont
  {San-Jose}, \citenamefont {Schiefele}, \citenamefont {Sturla},\ and\
  \citenamefont {Vozmediano}}]{amorim_2016}%
  \BibitemOpen
  \bibfield  {author} {\bibinfo {author} {\bibfnamefont {B.}~\bibnamefont
  {Amorim}}, \bibinfo {author} {\bibfnamefont {A.}~\bibnamefont {Cortijo}},
  \bibinfo {author} {\bibfnamefont {F.}~\bibnamefont {de~Juan}}, \bibinfo
  {author} {\bibfnamefont {A.}~\bibnamefont {Grushin}}, \bibinfo {author}
  {\bibfnamefont {F.}~\bibnamefont {Guinea}}, \bibinfo {author} {\bibfnamefont
  {A.}~\bibnamefont {Gutiérrez-Rubio}}, \bibinfo {author} {\bibfnamefont
  {H.}~\bibnamefont {Ochoa}}, \bibinfo {author} {\bibfnamefont
  {V.}~\bibnamefont {Parente}}, \bibinfo {author} {\bibfnamefont
  {R.}~\bibnamefont {Roldán}}, \bibinfo {author} {\bibfnamefont
  {P.}~\bibnamefont {San-Jose}}, \bibinfo {author} {\bibfnamefont
  {J.}~\bibnamefont {Schiefele}}, \bibinfo {author} {\bibfnamefont
  {M.}~\bibnamefont {Sturla}}, \ and\ \bibinfo {author} {\bibfnamefont
  {M.}~\bibnamefont {Vozmediano}},\ }\href@noop {} {\bibfield  {journal}
  {\bibinfo  {journal} {Phys. Rep.}\ }\textbf {\bibinfo {volume} {617}},\
  \bibinfo {pages} {1 -- 54} (\bibinfo {year} {2016})}\BibitemShut {NoStop}%
\end{thebibliography}
\end{document}